\documentstyle[aaspptwo]{article}
\begin{document}
\def\fe{$\langle{\rm Fe}\rangle$~}
\title{A new chemo-evolutionary population synthesis model for early-type galaxies. II: Observations and results}
\author{A. Vazdekis{$^{1}$ (asv@iac.es)}}
\author{R. F. Peletier{$^{2,1}$ (peletier@astro.rug.nl)}}
\author{J. E. Beckman{$^{1}$ (jeb@iac.es)}}
\author{E. Casuso{$^{1}$ (eca@iac.es)}}
\affil{$^{1}$Instituto de Astrofisica de Canarias,E-38200 La Laguna,
Tenerife, Spain} 
\affil{$^{2}$Kapteyn Instituut, Postbus 800, 9700 AV Groningen, The Netherlands}
\journalid{Vol}{Journ. Date}
\articleid{start page}{end page}
\paperid{manuscript id}
\cpright{type}{year}
\ccc{code}
\lefthead{Vazdekis, Peletier, Beckman \& Casuso}
\righthead{A new chemo-evolutionary population synthesis model for early-type galaxies}

\begin{abstract}
We present here the results of applying a new chemo-evolutionary stellar
population model developed by ourselves in a previous paper (Vazdekis {\it et
al.} 1996) to new high quality observational data of the nuclear regions of 
two representative elliptical galaxies and the bulge of the Sombrero galaxy. 
Here we fit in detail $\sim$20 absorption lines and 6 optical and 
near-infrared colors following two approaches: fitting a single-age single-metallicity 
model and fitting our full chemical evolutionary model. We find that all
of the iron lines are weaker than the best fitting models predict, indicating that the iron-abundance is anomalous and deficient. We also find that the Ca$_{I}$ index at $4227 {\rm \AA}$ is
much lower than predicted by the models. We can obtain good fits for all the
other lines and observed colors with models of old and metal-rich 
stellar populations, and can show that the observed radial gradients are due to metallicity decreasing outward. We find that good fits are obtained both with fully evolutionary models and with single-age  single-metallicity models. This is due to the fact that in the evolutionary model more than 80\% of stars form within 1.5~Gyr after the formation of the galaxies. The fact that slightly better fits are obtained with evolutionary models indicates 
these galaxies contain a small spread in metallicity.

\end{abstract}


\section{INTRODUCTION}
The study of the stellar populations and the distribution of metallicities 
plays an important role in our understanding of the star formation history 
of the galaxies. Their stellar populations are expected to be more complex than 
those of, for
example, globular clusters, which are thought to be composed of a single 
stellar population. In fact, Burstein {\it et al.} (1984) found differences 
when comparing colors and spectroscopic features of globular clusters with galaxies. Key parameters in the 
interpretation of the observed colors and the line-strengths are the 
metallicity and the age. The problem is that even in the simplest unresolved
 stellar systems their effects are very difficult to separate using only colors (O'Connell 1986, Renzini 1986, Buzzoni {\it et al.}
1992). 

Using colors together with absorption lines more accurate conclusions can be drawn. Although
every absorption line strength is dependent on different kinds of stars, in principle it should be possible to determine average metallicities or ages by carefully selecting features which more sensitive to the metallicity and others 
which are more sensitive to the age (e.g. Worthey {\it et al.} 1992). However the abundances of some elements may well evolve differently from those of others (e.g. $\alpha$-enhancement), and the conversion of ages and metallicities 
through models to observed colors and indices may be not unique due to problems in e.g. stellar evolution theory. Finally, the large velocity broadening in giant elliptical galaxies implies that only the strongest lines can be used to obtain physical information from their spectra.

In the process of understanding the stellar population of early-type galaxies we first developed a new spectrophotometric model, which can be used to interpret
observed colors and absorption lines of galaxies (Vazdekis {\it et al.} 1996,
 hereafter Paper I). The model is based on the latest improvements in stellar evolution theory and on the most recent stellar libraries. Instead of studying a large sample of galaxies using a few lines indices, as has been done before (e.g. Worthey {\it et al.} 1992, Gonzalez 1993), we preferred to obtain high quality observations of three representative early-type galaxies (two giant 
ellipticals and the bulge of the Sombrero galaxy), but in many colors and absorption lines, and to make very detailed fits to each index, to understand better global ages and metallicities, and also to follow the abundances of individual elements. Such analysis now is possible, since we could
calibrate our observations using the large sample of stars from the extended Lick-system (Worthey {\it et al.} (1994) hereafter WFGB).

In this paper we have applied our spectrophotometric population synthesis model following both the single-age single-metallicity and the chemical evolutionary approaches. We address here the problem of whether the conclusions we obtain 
depend on the stellar population synthesis method we use. In the end we find that the use of many indices does yield interesting information, and we show that we can learn more than by using only a few indices, as has been done in the past. At the same time we study the stellar population gradients in
the three galaxies.

This paper is organized as follows: in Section~2 we 
explain our observations and the method we use to derive the line-strengths. 
In Section~3 we fit our population synthesis model and discuss the results obtained by fitting the data. Finally in Section~4 we present our conclusions. 

\section{Observations and data reduction}
\subsection{Observations}
Long-slit spectra of three well know early-type galaxies were obtained with the
ISIS spectrograph on the 4.2m WHT in March, 1995, at the Observatorio del Roque 
de los Muchachos, La Palma. The spectra were taken using both arms of the 
instrument, with a large format TEK  windowed to
$1124\times600$ pixels, each with a size 
of $24~\mu m$ CCD chip. In the {\em blue} we used a grating of $600~lines/mm$
giving a sampling of $0.79~{\rm \AA pix^{-1}}$ while in the {\em red} we used a 
$300~lines/mm$ grating giving a sampling of $1.46~{\rm \AA pix^{-1}}$. Our spectra 
were taken in the range $3700-6300~{\rm \AA}$. This configuration allowed us to
cover almost the whole set of the absorption features contained in WFGB in
addition to some UV features as defined in Pickles (1985). 
In fact, in the blue arm we covered the range 3700 to 
4500~${\rm \AA}$, while in the red arm we covered the range 4800 to 6300~{\rm
$\AA$}. We could not observe the
Ca4455, Fe4531 and Fe4668 features because they fall in the crossover region of the dichroic. The measured resolution was $\sim 3.4~{\rm \AA}$ in the blue and 
$\sim 6.5~{\rm \AA}$ in the red spectra. We also lost the TiO$_{2}$ since this index
falls at the limit of the range covered. A set of stars of the sample of WFGB
was also observed to calibrate our line-strength measurements. We positioned the
slit on the major axis for NGC~4472, at $123^{\circ}$ for NGC~3379 (the major
axis is at $\sim 70^{\circ}$, see Peletier {\it et al.} 1990a) and on the minor
axis for the Sombrero galaxy. The exposure times were $1800~s$ for both frames.

\subsection{Data reduction}
All data reduction was done with the IRAF software package. The first step was 
subtracting the {\em bias} value calculated from the unilluminated portion of 
each frame. After this we flat-fielded using Tungsten lamp exposures. Next, the
spectra were wavelength calibrated using CuAr+CuNe calibration lamp exposures.
The obtained pixel scale was $57.535$ $km s^{-1}{\rm pix^{-1}}$ for the blue and
$79.926$ $km s^{-1} {\rm pix^{-1}}$ for the red spectra.
The following step was the {\em sky subtraction}, for which the outer parts 
of each galaxy frame (chosen at approximately $2 '$ from the center of the 
galaxy) were averaged to produce a mean sky spectrum, which then was 
subtracted from each frame. The last step is the elimination of pixels
affected by {\em cosmic rays} in each frame.

\subsection{The line-strength measurements}
We need a well-defined way to assign values to the strengths of features at
different radii in the galaxy. For this purpose we used the expanded 
Lick-system (WFGB). Here some indices (the atomic features) are
defined as equivalent widths and some (the molecular bands) as ratios of
line-depth to continuum in magnitudes and we have maintained these
definitions. To measure these line-strength indices along the major axis we
had to de-redshift each feature and the continuum on each side of it
using the recession velocity corresponding to each spectrum. To calculate the
rotation curve the spectra at each radius were cross-correlated with the 
observed spectrum of a stellar velocity standard star which looks most like 
the galaxy (with spectral type K III, see Section~2.3.1). This method is described in Bottema
(1988), following the paper of Tonry \& Davis (1979). Then we calculated the 
indices by co-adding a sufficient number of spectra in the spatial direction, 
so that a satisfactory continuum level was reached across the whole 
wavelength range. More details about the indices can be found in WFGB and
Burstein {\it et al.} (1984).

\begin{table}
\footnotesize
\begin{center}
\begin{tabular}{l|cc}
\hline\hline
\multicolumn{3}{c}{Correction factors}\\
\hline
Index&Vel. Disp.&Conv. to Lick\\
\hline\hline
UV CN	   &1.002&-\\
Ca H+K     &1.016&-\\
FeI+CN     &1.030&-\\
CN1        &1.036&-0.028\\
CN2        &1.046&-0.041\\
Ca4227     &1.334&-0.35\\
G-band     &1.032& 0.11\\
Fe4383     &1.074&-2.95\\
H$_{\beta}$&0.988&0.0\\
Fe5015     &1.132&0.37\\
Mg$_{1}$   &1.016&-0.058\\
Mg$_{2}$   &1.006&-0.053\\
Mg$b$      &1.106&0.16\\ 
Fe5270     &1.134&0.22\\ 
Fe5335     &1.253&0.10\\ 
Fe5406     &1.234&-0.05\\ 
Fe5709     &1.129&0.0\\ 
Fe5782     &1.242&0.0\\
NaD        &1.068&0.29\\ 
TiO$_{1}$  &1.067&0.006\\
\hline
\end{tabular}
\end{center}
\caption{Correction factors for velocity
dispersion (the given values correspond to NGC~4472 at 5'' calculated in EW)
and for the conversion to the expanded Lick-system (in {\rm \AA} except for CN1 CN2, Mg$_{1}$, Mg$_{2}$ and TiO$_{1}$ indices for which are given in magnitudes). 
The velocity dispersion correction factors are defined in Section 2.3.2, while the conversions to the Lick-system are constant quantities to be added to the measured indices.}
\end{table}

\begin{table}
\footnotesize
\begin{center}
\begin{tabular}{l|cccc}
\hline\hline
\multicolumn{5}{c}{Errors}\\
\hline
Index      &Poisson&Rot. Curve&Vel. Disp.&Conv. Lick\\
\hline\hline
UV CN	   &0.87 &0.35   &0.03 &-\\
Ca H+K     &0.44 &0.06   &0.02 &-\\
FeI+CN     &0.45 &0.15   &0.05 &-\\
CN1        &0.005&0.001  &0.001&0.017\\
CN2        &0.005&0.001  &0.001&0.017\\
Ca4227     &0.11 &0.06   &0.04 &0.23\\
G-band     &0.15 &0.09   &0.03 &0.30\\
Fe4383     &0.25 &0.09   &0.0  &0.46\\
H$_{\beta}$&0.17 &0.06   &0.0  &0.16\\
Fe5015     &0.36 &0.30   &0.02 &0.33\\
Mg$_{1}$   &0.003&0.002  &0.0  &0.005\\
Mg$_{2}$   &0.002&0.003  &0.001&0.006\\ 
Mg$b$      &0.17 &0.09   &0.06 &0.17\\ 
Fe5270     &0.17 &0.08   &0.02 &0.20\\ 
Fe5335     &0.18 &0.08   &0.02 &0.21\\ 
Fe5406     &0.15 &0.03   &0.01 &0.14\\ 
Fe5709     &0.09 &0.08   &0.01 &0.13\\ 
Fe5782     &0.09 &0.15   &0.0  &0.15\\ 
NaD        &0.11 &0.22   &0.05 &0.21\\ 
TiO$_{1}$  &0.003&0.001  &0.001&0.005\\
\hline
\end{tabular}
\end{center}
\caption{Uncertaintities in the indices due to photon statistics (Poisson), the selected
rotation curve zero-point, the velocity dispersion and the conversion to the 
extended Lick-system. The given photon error is an average value calculated 
at $\sim5$\arcsec~ from the center of the three galaxies.}
\end{table}

\begin{figure}
\plotone{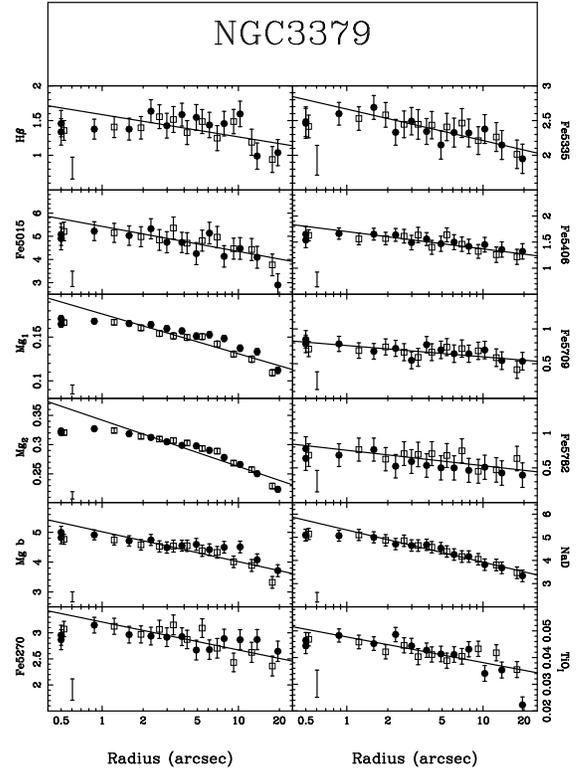}
\caption{Plot of the different indices of NGC~3379 obtained for a position angle
of $123^{\circ}$. Filled and open symbols indicate the values at each side of
the center respectively. We also have included linear fits (see also Table~3).
The error bars given here include all errors discussed in the text except the conversion to the extended Lick-system, which is given in the lower left corner.}
\end{figure}

\begin{figure}
\plotone{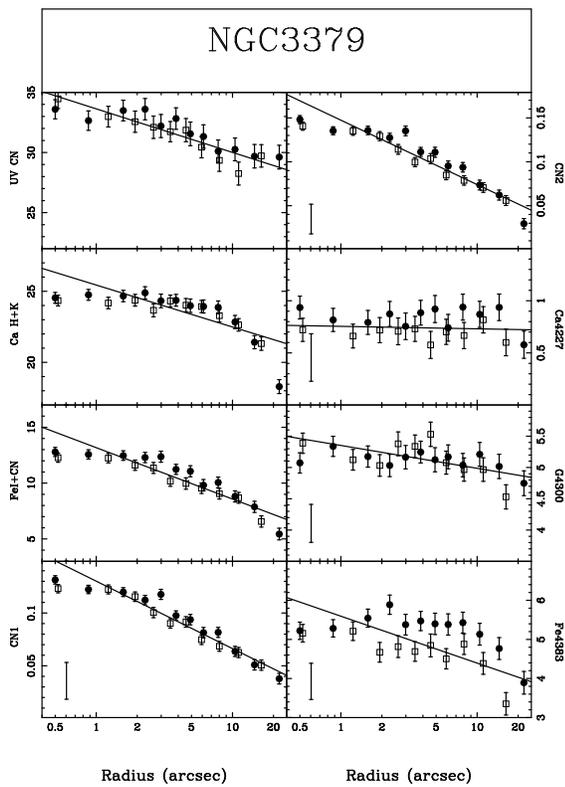}
\caption{The indices of NGC~3379 obtained in the UV region.}
\end{figure}

\begin{figure}
\plotone{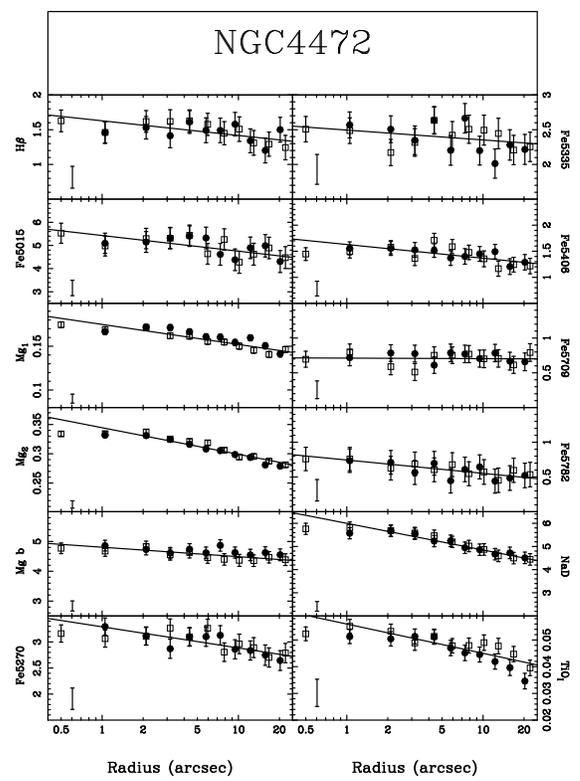}
\caption{The visible indices along the major axis of NGC~4472.}
\end{figure}

\begin{figure}
\plotone{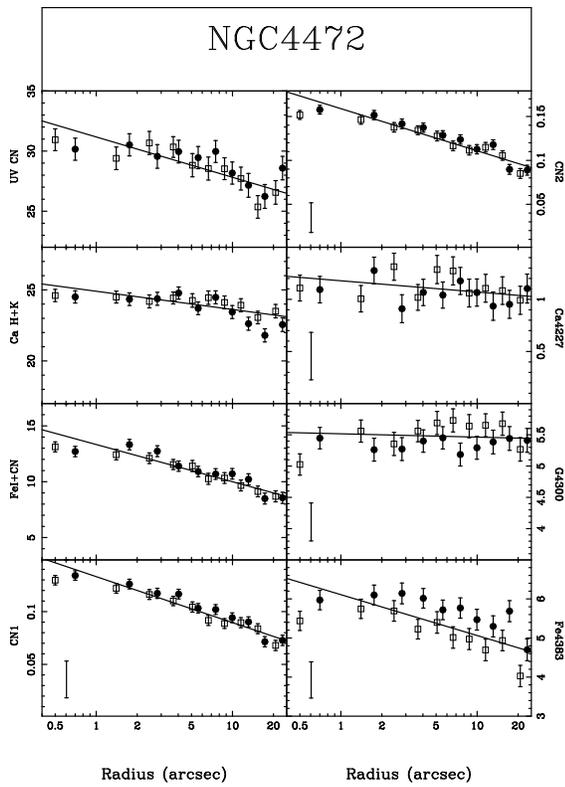}
\caption{The UV indices of NGC~4472.}
\end{figure}

\begin{figure}
\plotone{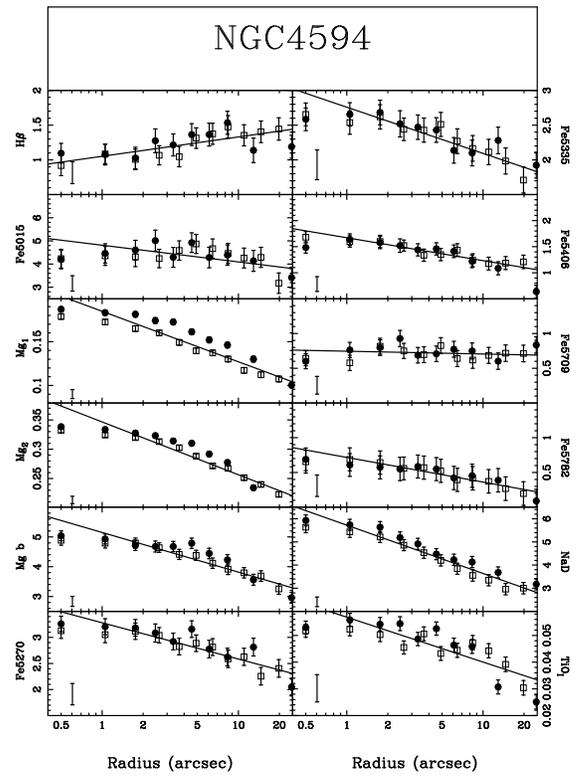}
\caption{The visible indices along the minor axis of Sombrero galaxy.}
\end{figure}

\begin{figure}
\plotone{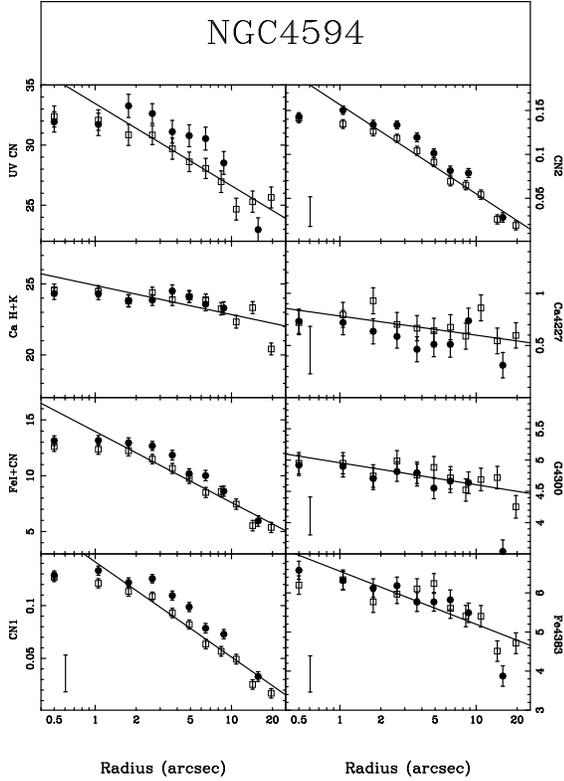}
\caption{The UV indices of the Sombrero galaxy.}
\end{figure}

\begin{table}
\voffset-3cm
\scriptsize
\begin{center}
\begin{tabular}{l|cccc}
\hline\hline
\multicolumn{5}{c}{NGC~3379}\\\hline                       
Index& a & $\epsilon$(a) & b & $\epsilon$(b) \\\hline
UV CN       & 33.6322 &  0.2918 & -3.6227 &  0.3837\\
Ca H+K      & 25.4420 &  0.3357 & -2.9515 &  0.4413\\
FeI+CN      & 13.1787 &  0.2556 & -4.6013 &  0.3229\\
CN1         &  0.1302 &  0.0023 & -0.0639 &  0.0029\\
CN2         &  0.1474 &  0.0032 & -0.0733 &  0.0040\\
Ca4227      &  0.7547 &  0.0411 & -0.0238 &  0.0519\\
G-band      &  5.3519 &  0.0775 & -0.3650 &  0.1019\\
Fe4383      &  5.5943 &  0.2005 & -1.2002 &  0.2532\\
H$\beta$    &  1.5859 &  0.0597 & -0.3210 &  0.0746\\
Fe5015      &  5.4273 &  0.1132 & -1.0814 &  0.1454\\
Mg$_{1}$    &  0.1765 &  0.0025 & -0.0455 &  0.0032\\
Mg$_{2}$    &  0.3419 &  0.0038 & -0.0791 &  0.0047\\
Mg$b$       &  5.0203 &  0.0822 & -1.0044 &  0.1029\\
Fe5270      &  3.2075 &  0.0583 & -0.5363 &  0.0729\\
Fe5335      &  2.6646 &  0.0394 & -0.4522 &  0.0493\\
Fe5406      &  1.7023 &  0.0308 & -0.3325 &  0.0385\\
Fe5709      &  0.7610 &  0.0277 & -0.1625 &  0.0346\\
Fe5782      &  0.7529 &  0.0364 & -0.2218 &  0.0455\\
NaD         &  5.3155 &  0.0434 & -1.3884 &  0.0542\\
TiO$_{1}$   &  0.0477 &  0.0012 & -0.0096 &  0.0015\\
\hline
\multicolumn{5}{c}{NGC~4472}\\\hline
Index& a & $\epsilon$(a) & b & $\epsilon$(b) \\\hline
UV CN       & 31.1736 &  0.4250 & -3.3511 &  0.4730\\
Ca H+K      & 24.9050 &  0.2165 & -1.2801 &  0.2484\\
FeI+CN      & 13.3326 &  0.1865 & -3.3352 &  0.2076\\
CN1         &  0.1332 &  0.0020 & -0.0436 &  0.0022\\
CN2         &  0.1589 &  0.0025 & -0.0480 &  0.0028\\
Ca4227      &  1.1783 &  0.0530 & -0.1077 &  0.0590\\
G-band      &  5.5165 &  0.0811 & -0.0531 &  0.0903\\
Fe4383      &  6.1080 &  0.1604 & -1.0448 &  0.1839\\
H$\beta$    &  1.6280 &  0.0441 & -0.2115 &  0.0491\\
Fe5015      &  5.4281 &  0.1392 & -0.6750 &  0.1550\\
Mg$_{1}$    &  0.1749 &  0.0018 & -0.0229 &  0.0020\\
Mg$_{2}$    &  0.3445 &  0.0022 & -0.0454 &  0.0025\\
Mg$b$       &  4.8062 &  0.0524 & -0.3174 &  0.0583\\
Fe5270      &  3.2897 &  0.0595 & -0.4047 &  0.0662\\
Fe5335      &  2.4941 &  0.0664 & -0.1417 &  0.0761\\
Fe5406      &  1.6290 &  0.0462 & -0.2629 &  0.0514\\
Fe5709      &  0.7112 &  0.0346 & -0.0079 &  0.0385\\
Fe5782      &  0.7444 &  0.0267 & -0.1918 &  0.0298\\
NaD         &  5.9980 &  0.0540 & -1.1499 &  0.0601\\
TiO$_{1}$   &  0.0560 &  0.0011 & -0.0109 &  0.0012\\
\hline
\multicolumn{5}{c}{NGC~4594}\\\hline
Index& a & $\epsilon$(a) & b & $\epsilon$(b) \\\hline
UV CN       & 33.4517 &  0.4922 & -6.8335 &  0.6023\\
Ca H+K      & 24.8937 &  0.2858 & -2.0476 &  0.3734\\
FeI+CN      & 13.9587 &  0.2868 & -6.3573 &  0.3510\\
CN1         &  0.1414 &  0.0040 & -0.0901 &  0.0049\\
CN2         &  0.1566 &  0.0040 & -0.1015 &  0.0049\\
Ca4227      &  0.7825 &  0.0477 & -0.1832 &  0.0584\\
G-band      &  4.9575 &  0.0521 & -0.3513 &  0.0680\\
Fe4383      &  6.5563 &  0.1437 & -1.3583 &  0.1831\\
H$\beta$    &  1.0519 &  0.0456 &  0.2782 &  0.0545\\
Fe5015      &  4.8142 &  0.1603 & -0.7205 &  0.1916\\
Mg$_{1}$    &  0.1847 &  0.0030 & -0.0577 &  0.0036\\
Mg$_{2}$    &  0.3466 &  0.0040 & -0.0903 &  0.0048\\
Mg$b$       &  5.1513 &  0.0762 & -1.3318 &  0.0910\\
Fe5270      &  3.2815 &  0.0520 & -0.6962 &  0.0621\\
Fe5335      &  2.7584 &  0.0490 & -0.6646 &  0.0586\\
Fe5406      &  1.6711 &  0.0284 & -0.4386 &  0.0351\\
Fe5709      &  0.7455 &  0.0329 & -0.0401 &  0.0394\\
Fe5782      &  0.7162 &  0.0195 & -0.3548 &  0.0233\\
NaD         &  5.7204 &  0.0832 & -2.0836 &  0.0994\\
TiO$_{1}$   &  0.0568 &  0.0016 & -0.0168 &  0.0019\\
\hline
\end{tabular}
\end{center}
\caption{Line-strength gradients along the major axis of NGC~3379 and NGC~4472
and along the minor axis of NGC~4594. The line-strength is given by:
$a+b\times\log(r)$ where r is in $arcsec$. $\epsilon$ represents the formal
errors.}
\end{table}

\subsubsection{The conversion to the expanded Lick-system}
Since we want to compare our data with the spectrophotometric model we developed 
in Paper I, which is based on the Lick-system, we need to transform our results 
to that system. Among the problems encountered in achieving this are the fact
that the instrument we used (ISIS) has a different spectral response, the fact 
that WFGB did not flux-calibrate their stars and the higher resolution of our data 
($\sigma\sim125kms^{-1}$ in the blue spectra and $\sim145kms^{-1}$ in the red spectra) compared to 
theirs ($\sim200kms^{-1}$). While the effect of a different instrumental 
response is almost negligible in the narrow indices it is important in 
the broader ones such as Mg$_{2}$ since both the index itself and the  
two pseudocontinua cover a wide range in wavelength. On the other hand the effect 
of having a higher resolution is important only in the narrower indices, which are affected 
in the same way as by the velocity dispersion broadening. Therefore, to transform our indices to the expanded Lick-system we first pre-broadened both spectra, the galaxy and the reference 
stars, so that to match the resolution of the Lick-system.
Next we compared the thus obtained line-strength measurements of our
stars with those given in WFGB to find an empirical average correction constant for 
each feature (see Table~1). In particular, 
the stars used for this conversion 
were HR~3461 (K0 III), HR~4521 (K3 III) and HR~4932 (G8 III) (see for details WFGB). 
We attribute the large correction factor found for the Fe4383
feature to the fact that its right-hand pseudocontinuum falls at the edge of our 
observed spectra, entering the cross-over wavelength region of the dichroic used 
in the observations. The large correction constant of Ca4227 is mainly due to the fact 
that this line is weak and the number of stars used for the conversion 
is small. As expected, we see that the most important corrections 
constants are those obtained for the molecular features.

\subsubsection{Correction for velocity dispersion of the galaxies}
To correct the measured indices for instrumental resolution and velocity
dispersion of the galaxy we used the spectra of a few K giants of the sample
of WFGB to calibrate their effects on the galaxy indices in the same way as
has been done before by e.g. Davies {\it et al.} (1993). An auto-correlation 
of the central spectrum of each galaxy gave its corresponding broadening, 
which includes both the instrumental and the real velocity dispersion. After this 
we convolved the pre-broadened stellar spectrum with a Gaussian of width $\sigma_{D}$ 
calculated to match that observed in the galaxy. The velocity dispersion of 
this Gaussian was (in pixels):
\begin{equation}
\sigma_{D}={\sqrt{\sigma_{G}^{2}-\sigma_{*}^{2}} \over \Delta v}
\end{equation}
where $\sigma_{G}$ is the observed velocity dispersion of the galaxy,
$\sigma_{*}$ is the measured instrumental profile (obtained by the 
cross-correlation of the star with itself) and 
$\Delta v$ is the conversion factor in $kms^{-1}{\rm pix^{-1}}$ (see
Section~2.2). For each index, {\it i}, an empirical correction factor,
$C_{i}(\sigma)$, defined as $i(\sigma)/i(0)$, was determined for all these
stars. We performed this step by calculating all the indices in EW and never 
in $magnitudes$. The resulting $C_{i}(\sigma)$ was found by taking the mean 
of the measured correction factors (see Table~1). From a quick look at these factors
we see that, in general, the higher the resolution and weaker the feature the
higher the correction. For example, the weakest line in the blue
spectra, the Ca4227, shows the highest correction (see also Fig.~13), while in
the red spectra this is the case for some of the iron features. 

Finally, the calculated velocity dispersions amounted to $250~kms^{-1}$ 
for NGC 3379, $320~kms^{-1}$ for NGC 4472 and $280~kms^{-1}$ for the
Sombrero. These values are not very different from those obtained by Davies
{\it et al.} (1983) who obtained $231~kms^{-1}$ for NGC~3379 and $310~kms^{-1}$ 
for NGC~4472 if we take into account that our error is about $\sim15~kms^{-1}$.
For the Sombrero galaxy Kormendy (1988) obtained $250~kms^{-1}$.
We attribute the differences to the moderate seeing we had during the observations.

\subsubsection{The resulting line-strengths}
The line-strength measurements for the three galaxies are shown in Figs.~1 to
6. In the three galaxies we see appreciable gradients for most of the
indices. However we detect only weak gradients for H$\beta$, the G-band and
the weakest lines: Ca4227, Fe5709 and Fe5782. The low slopes in H$\beta$ were
found useful for constraining the galaxy formation scenarios by Fisher {\it et
al.} (1995). In the next sections we will concentrate on the study
on the nuclear regions of these galaxies as well as on their gradients. 
To study the inner regions we have selected values for the indices corresponding to 5\arcsec~ 
from the center, and to study radial gradients we also selected values at 15\arcsec. We did not
go further inwards because our seeing was poor ($\sim$3\arcsec), to avoid possible nuclear 
emission lines (e.g. Goudfrooij \& Emsellem 1996, Boroson \& Thompson 1991), and because not much surface photometry is available in the inner regions. 

In Fig.~7 we give a comparison with the data from the current literature for
NGC~4472. Our Mg$_{2}$ index is slightly lower than the values of the other 
authors but we are in a better agreement with the data of Saglia {\it et al.} 
(1993). For H$\beta$ the agreement is generally good while our  
\fe values fall in
the middle range of other observations. For the sake of brevity, we have not shown other comparisons but, for example, for NGC~3379 our Mg$_{2}$ and H$\beta$ indices obtained are in a very good
agreement with Davies {\it et al.} (1993), while their \fe is higher than ours.

\begin{figure}
\plotone{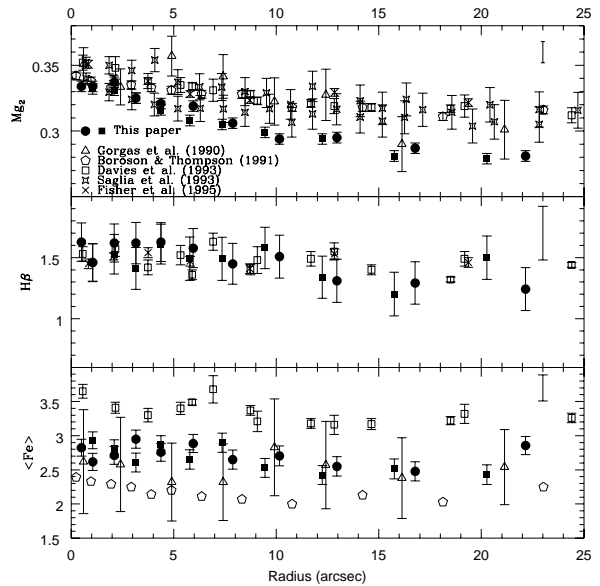}
\caption{We plot here for the standard giant elliptical NGC~4472 a comparison 
between our data (filled squares and circles indicating each side of the 
center) with recent data of other authors. The \fe index has been defined to be the average
between the Fe5270 and Fe5335 absorption lines. The Mg$_{2}$ data of the different
authors given here were projected to the major axis of the galaxy by Gonzalez 
\& Gorgas (1996) and were generously provided to us by Javier Gorgas. At a radius of 23\arcsec~ we indicate the error-bar due to the conversion to the extended Lick-system.}
\end{figure}
  
\subsection{Determination of the errors}
The main sources of errors are: the Poisson noise, the error in the adopted
zero-point of the calculated rotation curve (used to de-redshift the spectra),
the chosen velocity dispersion correction factors and the transformation to the
extended Lick-system.  

A quantitative estimate of the photon counting statistics has been 
carried out following the error analysis of spectroscopic features by Rich
(1988). The error in equivalent width is
\begin{equation}
\sigma(W)={\Delta \lambda N_{1} \over \overline{c}}\left [{1 \over N_{1}} +
({\sigma_{\overline{c}} \over \overline{c}})^{2}\right ]^{1/2}
\end{equation}
where $\Delta \lambda$ is the dispersion, $\sigma_{\overline{c}}$ is the error
in fixing the continuum, $N_{1}$ is the total number of counts in the line
bandpass (including negative counts), and $\overline{c}$ is the mean continuum
at the feature, defined as the value of the continuum point interpolated 
between the two continuum bands, at either side of the feature. The $\sigma(W)$
translates to a magnitude error $\sigma(m)$:

\begin{equation}
\sigma(m)=(-2.5 \log_{10} e)\left[ {1 \over N_{1}} + ({\sigma_{\overline{c}} 
\over \overline{c}})^{2}\right] ^{1/2}
\end{equation}

The uncertainty caused by the adopted zero-point of the rotation curve was 
estimated by calculating the rotation curves separately for each star and then
comparing the differences in the obtained line-strengths. The errors due to the
correction for velocity dispersion and the conversion to the Lick-system were
estimated in the same way, by looking at the dispersion in the
line-strengths obtained from different stars. Typical errors of the
various types are tabulated in Table~2. Here one can notice 
that in general the conversion to the expanded Lick-system introduces the most important 
uncertainty. This error is mainly due to the low resolution of the Lick-data, 
and therefore we are assuming their uncertainties. In Figs.~1 to 6 and
in Fig.~7 this error is given separately from the others in the corners of the
plots.

\section{Fitting the early-type galaxies}
In this section we apply the spectrophotometric population synthesis model 
developed in Paper I, which was especially designed to study early-type
galaxies. Briefly, the model makes predictions for the optical and IR colors and
25 absorption line indices. It is based on the new theoretical isochrones of
Bertelli {\it et al.} (1994) (calculated with solar abundance ratios), 
but converted to the 
observational plane by using empirical calibrations of individual stars (for 
details see Paper I). To calculate line-strengths it uses the latest stellar
spectral libraries (mainly WFGB). The model calculates the properties 
of a stellar system, starting from a primordial gas cloud and calculating the 
chemical evolution in a way which broadly follows the mathematical formalism of 
Arimoto \& Yoshii (1986). The model can also be used to obtain the integrated
colors of single-age single-metallicity stellar population (SSP). In this work
we will use both schemes. First we will apply a SSP model and later the full
chemical evolutionary model. We took the colors of NGC~3379 and 
NGC~4472 from Peletier {\it et al.} (1990a) and Peletier {\it et al.} 
(1990b) while those for NGC~4594 are taken from Hes \& Peletier (1993). As was 
explained in Section~2.3.3, we took the colors and line-strengths of two different regions of the galaxies: at 5\arcsec~  and 15\arcsec~  from the center (see Table~4). All these observables were calculated by linear interpolation between the neighbouring points and averaging between the two sides, except for the Sombrero galaxy at 15\arcsec~  where we only took into account the observables of the dust free side.
For NGC~3379 we took into account the fact that our spectra were taken with the slit positioned at $\phi=123^{\circ}$. For example, a major axis position of 5\arcsec~ corresponds to a distance of 4.8\arcsec~  from the center along the slit.

\begin{table}
\footnotesize
\begin{center}
\begin{tabular}{c|cc}
\hline\hline
\multicolumn{3}{c}{The selected positions}\\
\hline
\multicolumn{3}{c}{NGC~3379 (V$_{r}$=889${\rm km s^{-1}}$,r$_{e}$=37.5\arcsec )}\\
\hline
arcsec&$Kpc$&Fraction r$_{e}$\\
\hline
5.0&0.29&0.13\\
15.0&0.86&0.40\\
\hline
\multicolumn{3}{c}{NGC~4472 (V$_{r}$=983${\rm km s^{-1}}$,$r_{e}$=114.0\arcsec)}\\
\hline
5.0&0.32&0.04\\
15.0&0.95&0.13\\
\hline
\multicolumn{3}{c}{NGC~4594 (V$_{r}$=1082${\rm km s^{-1}}$,$r_{e}$=61.8\arcsec)}\\
\hline
5.0&0.35&0.08\\
15.0&1.05&0.24\\
\hline
\end{tabular}
\end{center}
\caption{The selected positions (in $arcsec$) for the three galaxies. The distances from the center of the galaxies were calculated taken their recesion velocities from the {\em Third Reference Catalogue of Bright Galaxies} and H$_{0}=75~kms^{-1}Mpc^{-1}$. The effective radii $r_{e}$ of the ellipticals were taken from Burstein {\it et al.} (1987) as given by Peletier {\it et al.} (1990a) and for the Sombrero galaxy from Hes \& Peletier (1993).}
\end{table}

\subsection{Fitting with the single-age stellar population model}
Using the V-K - Mg$_{2}$ and B-V - Mg$_{2}$ diagrams we showed in Paper I that
to fit this set of galaxies solar metallicities or larger values are required. 
In Fig.~8 we plot a number of color-color diagrams. In Fig.~9 we have selected two key colors, $B-V$ and $V-K$, and plotted a representative feature of each element versus these colors. Finally, in Fig.~10 we have plotted various index-index diagrams and since the number of features is large we selected as references for these plots three of the most commonly used indices in the literature: H$_{\beta}$, Mg$_{2}$ and \fe. To
obtain these figures we used the two forms of the IMF defined in Paper I: the 
{\em unimodal IMF} with a power law of slope $\mu$ as a free parameter (where
1.35 corresponds to the Salpeter value), and a {\em bimodal IMF}, which is equal to the unimodal IMF
above $0.6~M_{\odot}$, but reduces the influence of stars with masses below 
$0.6~M_{\odot}$. The two IMF's used here have a lower mass-cutoff of 0.1~M$_{\odot}$ and an upper mass-cutoff of 72~M$_{\odot}$. Looking at the respective diagrams in these figures we see that a difference 
between models using the two IMF's is seen only in the redder spectral indices such as TiO$_{1}$ or the redder colors such as $V-K$. 
Of course, the unimodal IMF gives us higher line-strengths since the number 
of low-mass stars is higher and therefore the relative number of {\em diluting} 
blue stars is lower. 

Looking at all these figures, but excluding those diagrams 
that contain iron features, we infer that to fit this set of galaxies we 
need either solar metallicity and very high ages (say 15~Gyr or even more) or 
metallicities that are higher than solar and lower ages (around say 8~Gyr) but never very low ages. We also see that the colors seem to be best fitted using 
metallicities lower than those used for line-strengths as one can see when comparing the best fits of the 
color-color diagrams with those of the index-index diagrams. This shows 
how important is the combination of colors and indices in this kind of studies 
(see also Sections~3.1.4 and 3.1.6). From the index vs. H$_{\beta}$ diagram one can conclude that the observed gradients inside each galaxy must be attributed to the metallicity rather than age variations. From the index-index diagrams we see that while most of the observed indices can be easlily fitted when plotted versus Mg$_{2}$ or H$_{\beta}$, this is not the case when plotted versus the 
\fe index. In the following we will discuss various 
aspects of the fits.

\begin{figure}
\plotone{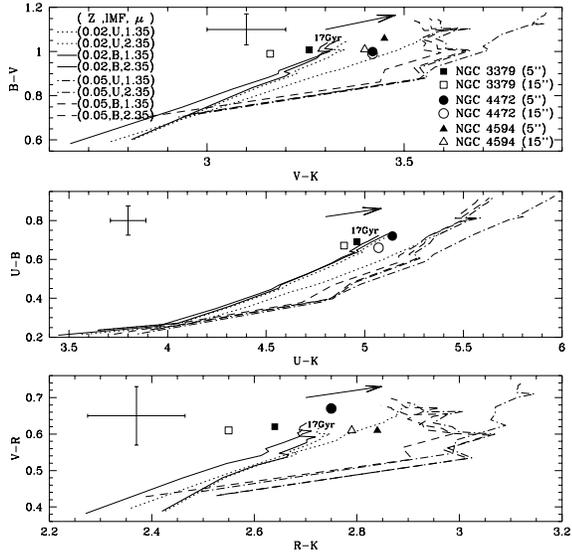}
\caption{Plots of different color-color diagrams using our SSP models. To obtain the synthetic colors we used unimodal (U) and bimodal (B) IMF's as defined in Paper I. Notice that the most important changes when using these IMF shapes are in the redder colors. We have varied the age from 1 to 17~Gyr, the metallicity and the slope of the IMF ($\mu=1.35$ and $\mu=2.35$). We have also plotted the observed colors for
the three galaxies analyzed here at different radii. Finally we have plotted characteristic errors (see Table~2) as well as the dust extinction vectors corresponding to $A_{V}=0.2 mag$ (following the simple reddening law of Rieke \& Lebofsky 1985).}
\end{figure}

\onecolumn
\begin{figure}
\plotone{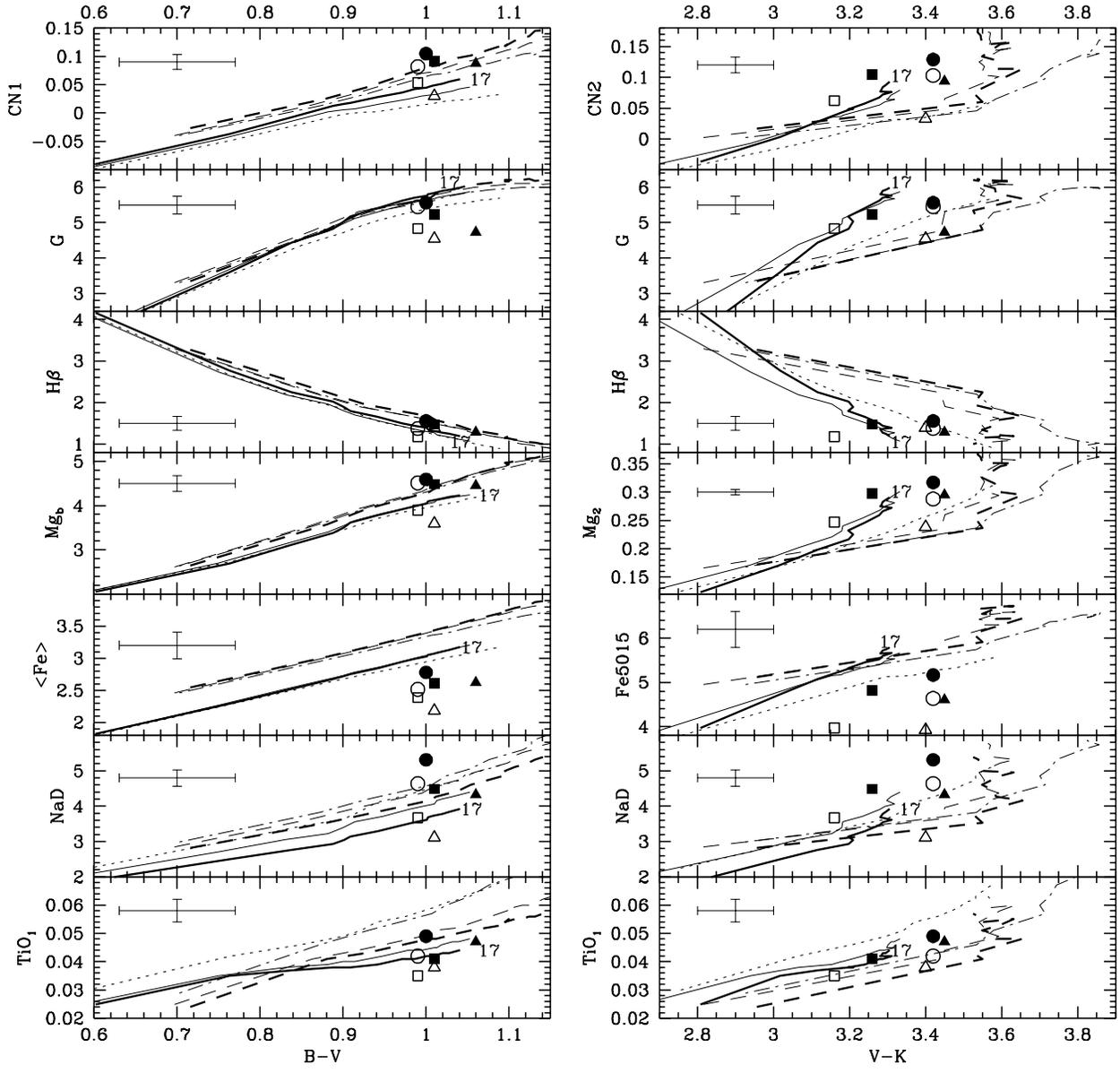}
\caption{Plot of different color-index diagrams. The symbols and models as in Fig~8. Compared to that figure we have not included 
models with unimodal IMF and $\mu=1.35$ since the synthetic observables are almost identical to the ones obtained with a bimodal IMF with the same slope.}
\end{figure}

\begin{figure}
\plotone{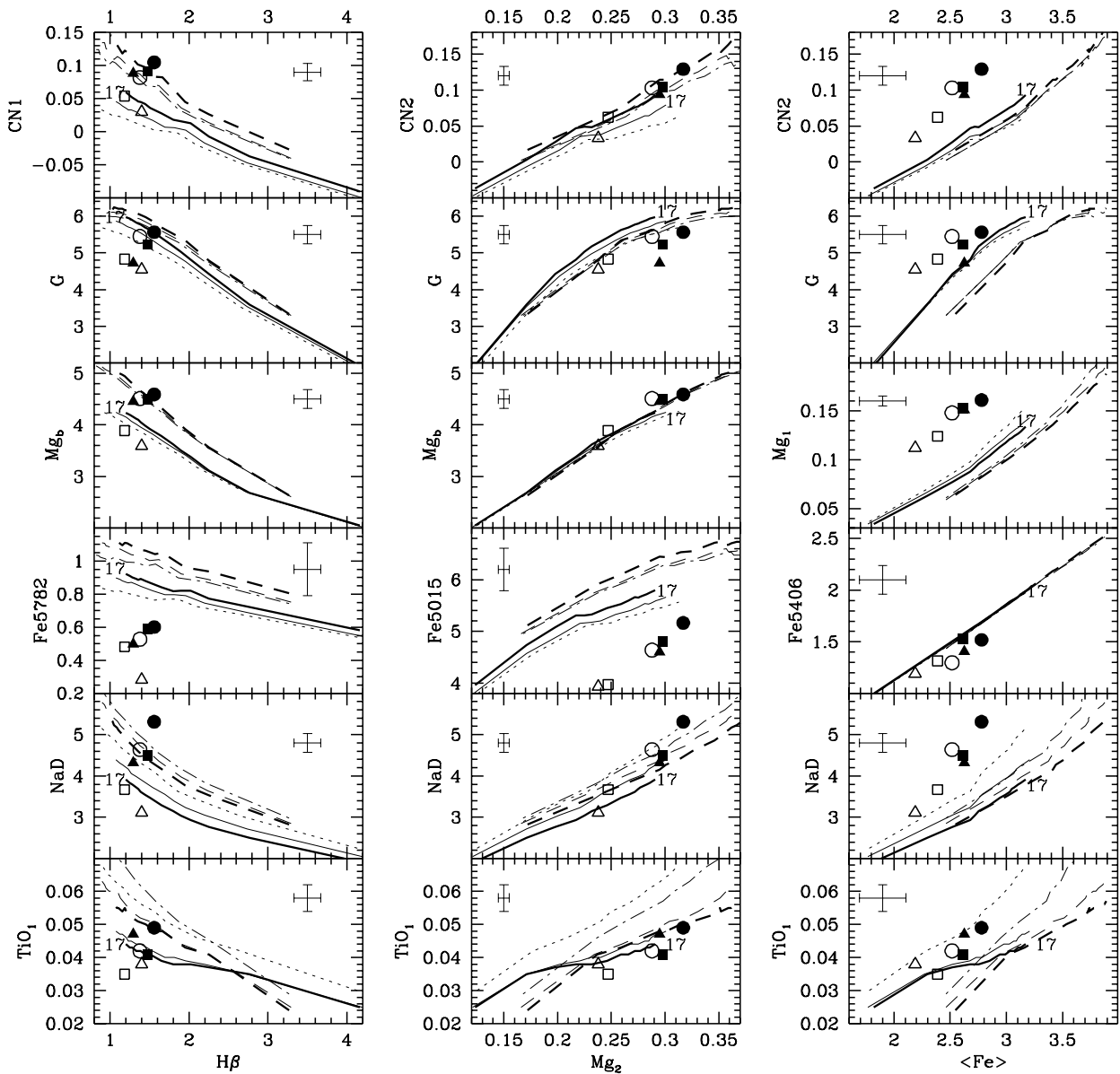}
\caption{The synthetic index-index diagrams compared with observations. The symbols and models as in Figs.~8 and 9. Here we also excluded models with a unimodal IMF and $\mu=1.35$.}
\end{figure}
\twocolumn

\subsubsection{The iron features}
In Fig.~9 we see that all the iron lines from the model are stronger than 
those observed, and in Fig.~10 we see that the models are always able to fit the different index-index plots, 
except those against \fe. This suggest that iron is anomalous and deficient compared to the other indices and colors. This shows that solar ratios are not adequate to fit this kind of galaxies, confirming the results of Peletier (1989), Worthey {\it et al.} (1992), Gonzalez (1993) and 
Davies {\it et al.} (1993). These results do not depend on whether we are 
working with a unique SSP or with a mixture of SSP's of different ages and/or 
metallicities (as in Section~3.2) as can be easily inferred from the different Fe vs. Mg$_{2}$ plots. To try to explain these results we applied some simple models based on the hypothesis of $\alpha$-enhancement. In reality, this phenomenon in principle affects all the parameters of stellar evolution and therefore a new set of isochrones ought to be obtained calculating observables
for integrated stellar populations. As a first 
approximation we follow the conclusions of Weiss {\it et al.} (1995) that for 
the calculation of $\alpha$-enhanced indices one can use the isochrones 
calculated for solar abundance ratios, keeping the global metallicity constant 
but changing the ratios. In Fig.~11 we find better solutions for the
Mg and Fe 
indices, but worse for Na (at least with the adopted ratios). For an IMF slope
of 1.35 it does not make any difference whether a unimodal or bimodal IMF is
chosen. From the Fe vs. Mg$_{2}$ plots we see that to fit the data we need 
[Mg/Fe] in the range 0.3 to 0.7, 
in agreement with the results obtained by Weiss 
{\it et al.} (1995). 
 This number does not seem to vary inside each galaxy. To conclude,
we find that the metallicity (in terms of Z) determined from the Fe lines
is different from that determined from the Mg lines.

\begin{figure}
\plotone{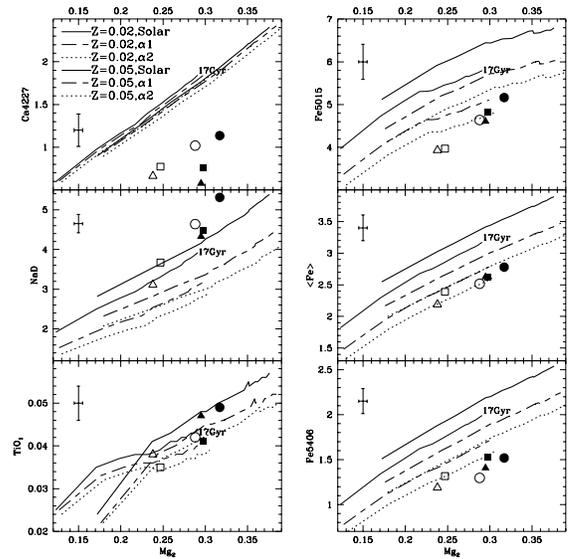}
\caption{$\alpha$-enhanced test diagram using the [Mg/Fe]=0.4 ($\alpha 1$) 
and [Mg/Fe]=0.6 ($\alpha 2$) enhanced mixtures of Weiss {\it et al.} (1995). 
The synthetic values were obtained for a bimodal IMF with slope 1.35. We plot here different representative features (including three
iron lines) vs. the Mg$_{2}$. These mixtures make our prediction for the iron 
features better but the NaD index worse.}
\end{figure}

\subsubsection{The Ca4227 feature}
We left the discussion of the Ca4227 line for a separate section, 
since this is the worst fitting line. In Fig.~12 we plot different representative 
features and the color $V-K$ versus this feature. Apart from our models we also plot those of Worthey 
(1994). We see here that the discrepancy between predictions and observations 
is more dramatic than was found for the iron lines. Independently of the IMF 
slope, age or metallicity, the Ca4227 line in the three galaxies is much lower
than predicted by the models. It has been suggested before by O'Connell (1976) and Worthey (1994) that calcium tracks iron. In fact in this Figure 
we see that the fits obtained when plotting this line vs.  \fe are not as bad as when plotting it vs. the other features. Even so this line cannot be fitted in an acceptable way. Neither is the result due to a deficiency in the
supporting library of stars, since for all the other indices (except those
dominated by Fe) reasonable fits can be achieved. It looks as if Ca is
depleted in this type of galaxies. If however we look at other Ca lines we 
cannot easily confirm this.  We could not observe the Ca4455 feature (a
blend containing an important contribution of Ca$_{I}$) as a result of the wavelength
position of the dichroic used for the observations, and the 
Ca II near-IR triplet seems to give only a small depletion.
Terlevich {\it et al.} (1990) for NGC 4472 found a total EW of $\sim 6.7$
for the two strongest lines, corrected for velocity dispersion. If we model this
feature, using the stellar library of D{\'{\i}}az {\it et al.} (1989) and 
preserving their definitions (for details see Paper I), we see that our best 
fitting models predict an EW of around $\sim 8.0$ (see Tables~7 and 8). This
means that we find some depletion, but much less than for the Ca4227 feature.

As can be seen in the Ca4227 vs. Mg$_{2}$ diagram of Fig.~11 
it is easy to understand that an $\alpha$-enhanced scenario does not fit the 
observed Ca4227 since this feature is almost all due to Calcium, another 
$\alpha$-element. The same plot also shows us that, even considering this 
scenario, the small contribution to this feature by Fe$_{I}$ (around $\sim 20\%$
by inspection in the Arcturus atlas) cannot explain the observed data. Since the
$^{48}Ca$ isotope is much less abundant than the $^{40}Ca$ we also cannot 
attribute this effect to the fact that $^{48}Ca$ isotope is not produced in 
quantities appropriate to its solar abundance and because it is made in Type
Ia supernovae (those that ignite a carbon deflagration very near the 
Chandrasekhar mass) as recently suggested by Woosley \& Weaver (1995).

Rose (1984) measured Ca$_{II}$ H and K indices in stars, finding that Hyades and Pleiades 
dwarfs present strong Ca$_{II}$-emision compared with field dwarfs. He 
suggested that many of the principal absorption features in the blue spectral
region are affected by stellar activity. However, in another paper
 (Rose 1994), he himself measured the Ca4227 line in 47~Tuc and M32, and reached the opposite
conclusion to that which we find here. This suggests that this effect, if it is 
real, will be present only in bright early-type galaxies in the same way as 
Mg tends to be enhanced with respect to Fe in this kind of galaxies. If 
this result is general it will have important consequences for our knowledge 
of element-synthesis in type II supernovae. 

However we cannot strongly affirm the anomalous behavior of Ca for two principal reasons. Firstly, 
Ca4227 is a weak line and secondly because, as one can see from Table~1, its conversion to the extended Lick-system entails one of the highest 
corrections (-0.35). The error in this correction is quite large, as a result of
which this depletion formally only is a $2.5\sigma$ result. To see whether the 
depletion is really present, we show the spectrum of NGC~4472 together with the
star HR~3461, for which WFGB measure an EW for the Ca4227 of $1.05{\rm \AA}$ and we measure $0.81{\rm \AA}$, while for the galaxy we find a value $1.14{\rm \AA}$ (the last two EW converted to the Lick-system). For the present we can say only that this will need more study and that observations of the
Ca$_{II}$ triplet feature in the near-IR may give additional help.

\begin{figure}
\plotone{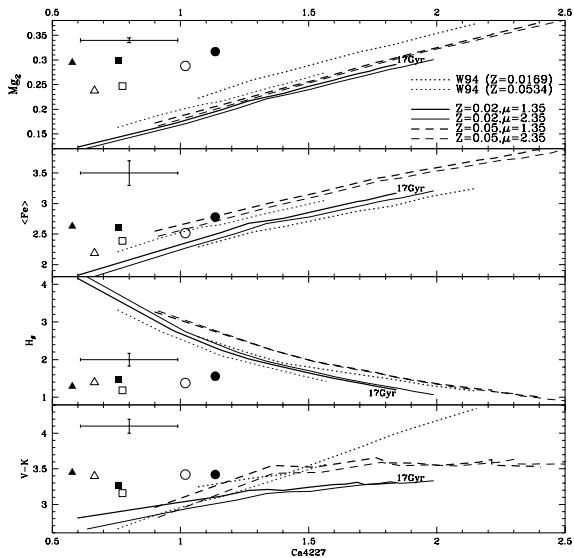}
\caption{Diagrams of the Ca4227 vs. Mg$_{2}$, \fe, H$\beta$ and the color $V-K$. Our synthetic
values were calculated using a bimodal IMF of slope 1.35 and 2.35 
for solar and 2.5 times solar metallicities, while the age was varied from 1 to
17~Gyr. Also shown are predictions of the model of Worthey (1994) (for Salpeter
IMF). Notice that for none other metallicities, ages or IMF slopes can we fit the
data.}
\end{figure}

\begin{figure}
\plotone{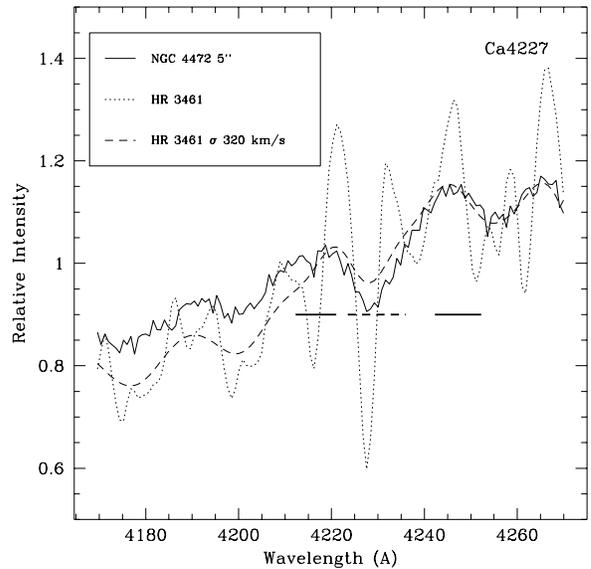}
\caption{The region around the Ca4227 index of the spectrum of NGC~4472, a template 
star of WFGB and the same star but convolved with a Gaussian to match the 
galaxy. The thick horizontal lines represent the regions covered by the 
line and the pseudocontinua. Notice how the pseudocontinua definition shows its inadequacy in the 
well-resolved spectrum.}
\end{figure}

\subsubsection{All the synthetic results combined into one value. 
The Merit function.}
Since the number of observables is large, we have implemented a 
{\em merit-function}, {\em M}, defined as

\begin{equation}
{\em M}=\sum_{i=1}^{n}H_{i} \left ( {{G_{i}-S_{i}} \over E_{{obs}_{i}}} 
\right )^{2}
\end{equation}
where $S_{i}$, $G_{i}$ are the synthetic and observed variable $i$ 
respectively and $E_{{obs}_{i}}$ is the corresponding observational error. All
these quantities must be expressed in magnitudes to perform the calculation of
the {\em M}. Finally, $H_{i}$ is the assigned weight. 

In practice we determine a separate merit figure for the set of colors $\em M_{c}$ and
for the set of line-strengths $\em M_{l}$. The merit function determines the 
goodness of a fit for all observables taken together. The problem of optimizing 
a fit is now reduced to finding a minimum for this merit function. Since the 
observational errors are non-zero the code calculates the maximum acceptable 
value for the merit, {\em M$_{max}$} using the same equation, but using the 
observational error instead of $G_{i}-S_{i}$. Any solution that gives a value 
of merit smaller than {\em M$_{max}$} is in principle acceptable. 

In this paper we have assigned the same total weight to the colors and to the 
lines. Therefore the final global merit is calculated as follows

\begin{equation}
{\em M}=M_{c}+M_{l} \left ( M_{{c}_{max}} \over M_{{l}_{max}} \right )
\end{equation}

\subsubsection{The best fits obtained with a unique SSP}
To find the best solutions we first have to give weights to each color and 
index. As has been shown before it is clear that whatever the metallicities,
ages or IMF slopes are we cannot fit the iron lines, possibly because of 
Fe deficiency. 
For that reason in this paper we give a weight of 0 to indices composed mainly
by iron. We also assign a weight of 0 to the Ca4227 aborption line due to its
apparently strong {\em depletion}. To the other observables we give 
a weight of 1 except for the CN1 and CN2, because they are the same
index (they only differ in the definition of the blue pseudocontinuum bandpass)
and therefore we assing a weight of 0.7 to each instead of 1, so that their global weight is higher
than 1 but lower than 2. The given weights are summarized in Table~5 (Case A). 
On the basis of the observational errors given in Table~2 (excluding the error from the conversion to the expanded Lick-system) and on the assigned 
weights, only merits smaller than 10.0 can be considered as acceptable fits. However, as one can see later, the obtained merits are often as high as 20. These numbers are probably still acceptable, since
we did not include the systematic uncertainties in converting our indices to the expanded Lick system and systematic errors in the theoretical models.

Using this merit function we scanned the 3-parameter space (Z,age,$\mu$) for 
each galaxy. Since our SSP models make use of the isochrones with a large step in 
metallicities (see Paper I) we made a new grid of synthetic values by interpolating 
linearly between the output obtained for Z=0.008, Z=0.02 and Z=0.05 to obtain the 
synthetic observables corresponding to Z=0.012, Z=0.016, Z=0.03 and Z=0.04. Fig.~14 shows the 
merit values obtained for NGC~3379 at the two selected positions 5\arcsec~  and 15\arcsec. 
To obtain this figure we used a unimodal IMF with slopes varying from 0 to 2.3, 
while the age was varied from 1 to 17~Gyr. 
We plot the contours and the grey-levels for merits which are lower than $2.5 \times M_{max}=25.0$. 
We choose this maximum acceptable limit to be safe and to avoid
excluding even moderately possible solutions. In Fig.~15 we scanned the same parameter space to fit this galaxy but using a bimodal IMF. Comparing the two figures it is clear that the unimodal IMF gives worse fits and for that reason
 we will work throughout this paper with a bimodal IMF. The same results were found in Paper I using our full chemo-evolutionary population synthesis model.
 The main difference that we found looking at the fits obtained
with the two IMF's, is the fact that the best fits obtained using a 
unimodal IMF require lower slopes than those obtained with the bimodal IMF. We expect this 
since the unimodal IMF yields a higher number of low-mass stars (see Paper I). 
As a general conclusion we can say that these galaxies cannot be fitted assuming low values for the age. The inner regions of the three galaxies must be metal-rich since merit figures lower than 25.0 are
only found for metallicities higher than solar. In particular NGC~4472 is even more metal-rich than the other two galaxies.

\subsubsection{Stellar Population Gradients} 

Looking at Figs.~14 and 15 we see that to fit the two selected positions of NGC~3379 we must keep the age almost constant with values around 13~Gyr, while the metallicity varies from values higher than solar (Z$\sim$0.03) at 5\arcsec~
  to lower than solar (Z$\sim$0.016) at 15\arcsec. We also notice that it is not possible to obtain acceptable fits if we maintain the metallicity constant but change the age and/or the IMF slope. However this is not the case for 
NGC~4472 (see Fig.~16) for which we find acceptable solutions if we decrease the age from 10~Gyr to 8~Gyr going outward and maintain the metallicity constant at $\sim$Z=0.04. Of course we also find a fit if we keep the age constant around 10~Gyr but decrease the metallicity from Z=0.04 to Z=0.03 going outward. For the bulge of the Sombrero galaxy (see Fig.~17) a decrease in the metallicity is 
required when going outward and no age-variable metallicity-constant solution will do. We see that this galaxy seems to behave in the same way as NGC~3379 but with ages slightly lower.  

We can conclude that
the observed radial index gradients are attributable to metallicity gradients rather than gradients in age.
The fact that this conclusion cannot be completely verified in NGC~4472 may be 
explained if we take into account that this galaxy is much larger than the other two, and thus the measured positions represent relatively small fractions of its effective radius (see Table~4). 

\paragraph{Testing for the presence of dust.}
To test for the presence of dust in these galaxies we have used the simple reddening law of Rieke \& Lebofsky (1985) and applied it to the synthetic colors using steps in $A_{B}$ of 0.1 $mag$. We scanned again the whole parameter space 
(Z,$\mu$,age) and examined the fits using the merit function. 
We found that the inclusion of dust does not improve the fits, 
except for the Sombrero galaxy at 15\arcsec~  where we obtained a better fit including $A_{B}=0.3 mag$.

\subsubsection{Mg overabundant or Fe deficient ?}
From the previous color-index and index-index diagrams we have inferred that the iron lines are not well fitted by the models. However to be more complete one might think to include them in the fitting procedure and neglect the magnesium features or take other weighting distributions to check the validity of our selection and for a better interpretation of our fits. For that purpose, 
together with the choice of weights used throughout this work (Case A) in which we have not given any weight to the Fe and Ca lines, we defined four alternative merit functions. In Case B we have taken into account all the iron features and assigned each a weight of 0.43, so that their global weight is 3, the same 
as the sum of the magnesium features. In case C 
instead of neglecting the iron lines as in Case A, we gave a weight of 0 to the three magnesium lines. In Case D we only took into account the iron lines and
H$_{\beta}$ and have neglected all the other lines. Finally, in Case E we 
neglected both the iron and magnesium as well as the Ca4227 line and gave a weight of 1 to each of the other features. Then, as an illustrative example, we used our SSP models with a bimodal IMF to fit NGC~4472 at 5\arcsec. In Table~6 we summarize the best fits obtained with these Merit functions, showing us that 
the required metallicities, ages and IMF slopes vary little amongst the Cases except for Case D where e.g. the metallicity obtained is much lower than for the others. The fact that the merit worsens considerably 
when combining Fe with other indices, and the fact that in Case D, where we excluded all the 
indices except the iron lines, a fit is obtained 
 which is not better than in Case A 
(where we kept most of the indices) encourages us to use Case A throughout this paper. From this excercise we can conclude that the iron lines yield worse fits and that these indices should not be used in combined fits with either colors or 
other indices, indicating that in this set of early-type galaxies the iron is in fact  {\it deficient}. Another conclusion is that the global metallicity inferred must depend on whether we use magnesium or iron lines as the prime indicators. This result also shows how important 
it is to use colors and line-strengths together, since in this case
the number of constraints is much higher than if we use colors alone.

\begin{table}
\footnotesize
\begin{center}
\begin{tabular}{l|ccccc}
\hline\hline
\multicolumn{6}{c}{Weights of the Merit Function}\\
\hline
&\multicolumn{5}{c}{Case}\\
Color/Index &A&B&C&D&E\\
\hline\hline
U-V        &1&1&1&1&1\\
B-V        &1&1&1&1&1\\
V-R        &1&1&1&1&1\\
V-I        &1&1&1&1&1\\
V-J        &1&1&1&1&1\\
V-K        &1&1&1&1&1\\
CN1        &0.70&0.70&0.70&0   &0.70\\
CN2        &0.70&0.70&0.70&0   &0.70\\
Ca4227     &0   &0   &0   &0   &0\\
G-band     &1   &1   &1   &0   &1\\
Fe4383     &0   &0.43&0.43&0.43&0\\
H$_{\beta}$&1   &1   &1   &1   &1\\
Fe5015     &0   &0.43&0.43&0.43&0\\
Mg$_{1}$   &1   &1   &0   &0   &0\\
Mg$_{2}$   &1   &1   &0   &0   &0\\ 
Mg$b$      &1   &1   &0   &0   &0\\ 
Fe5270     &0   &0.43&0.43&0.43&0\\ 
Fe5335     &0   &0.43&0.43&0.43&0\\ 
Fe5406     &0   &0.43&0.43&0.43&0\\ 
Fe5709     &0   &0.43&0.43&0.43&0\\ 
Fe5782     &0   &0.43&0.43&0.43&0\\ 
Na D       &1   &1   &1   &0   &1\\ 
TiO$_{I}$  &1   &1   &1   &0   &1\\
\hline
\end{tabular}
\end{center}
\caption{Five different cases for the assignation of the weights of the Merit 
Function. Case A is the distribution used throughout this paper.}  
\end{table}

\begin{table}
\footnotesize
\begin{center}
\begin{tabular}{c|ccc|c}
\hline\hline
\multicolumn{5}{c}{The best fits for NGC~4472 (5\arcsec)}\\
\hline
Case & Z &Age($Gyr$) &$\mu$ & Merit\\
\hline
A&0.04 &10 &2.3&16.4\\
B&0.03 &14 &2.3&30.9\\
C&0.04 &8 &2.3&37.1\\
D&0.016&12&2.3&17.8\\
E&0.05 &8 &2.3&17.8\\
\hline
\end{tabular}
\end{center}
\caption{The best merit values obtained for NGC~4472 at 5\arcsec~  using the different weight distributions tabulated in Table~5. Case A is the one used throughout this paper. We used here our SSP models with a bimodal IMF.}
\end{table}

\begin{figure}
\plotone{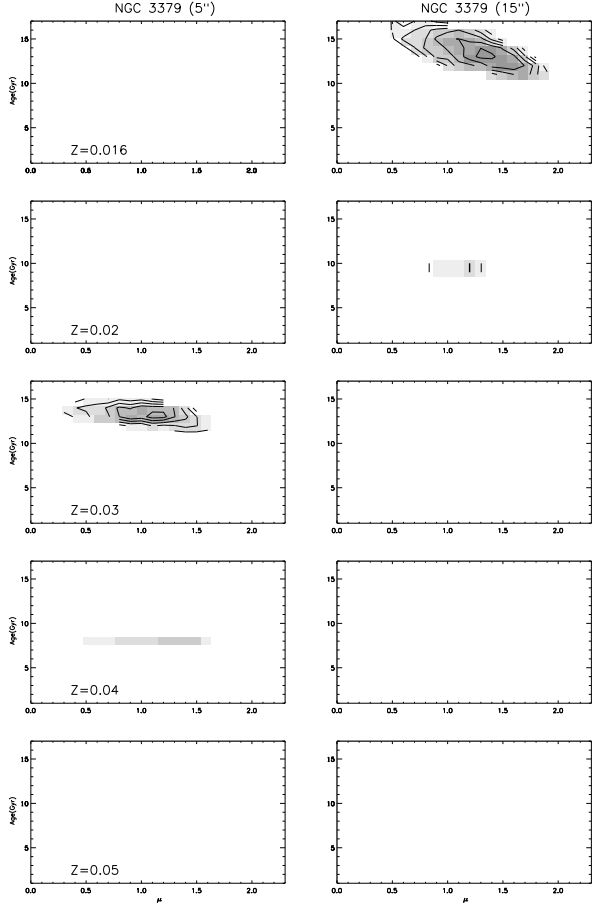}
\caption{The merit values obtained for NGC~3379 with our SSP models in the ($\mu$,Age) parameter space for different metallicities and using a unimodal IMF. The
contours are separated by steps of 1 from 10.0 to a maximum of 25.0. In the
grey-scale: black indicates merits of 10.0 or lower, while merits of 25.0 or higher are white. The estimated highest acceptable merit is 10.0 (calculated on 
the basis of the observational errors as explained in Section~3.1.4).}
\end{figure}

\begin{figure}
\plotone{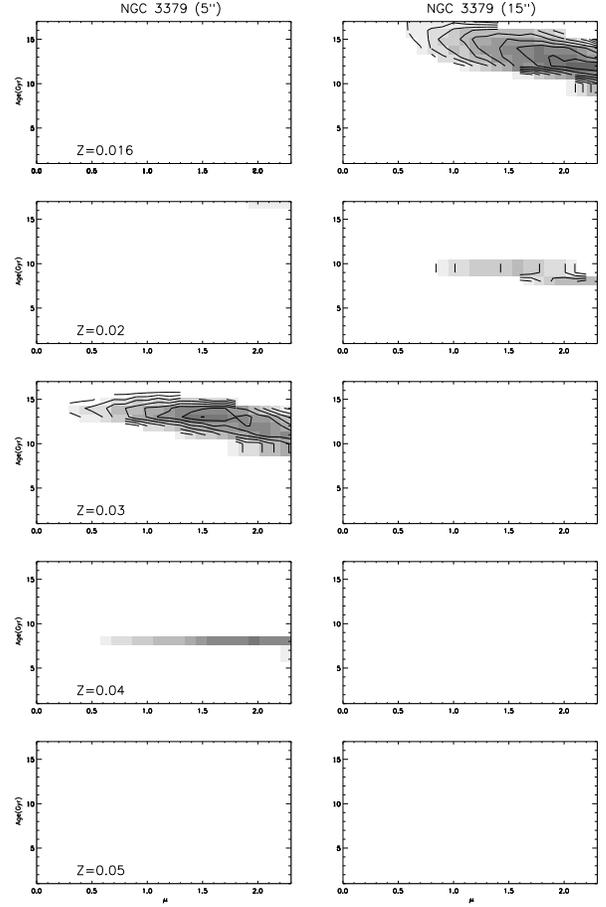}
\caption{The merit values obtained for NGC~3379 with our SSP models in the ($\mu$,Age) parameter space for different metallicities and using a bimodal IMF. Notice that the fits are better than those obtained with a unimodal IMF (see Fig.~14).}
\end{figure}

\begin{figure}
\plotone{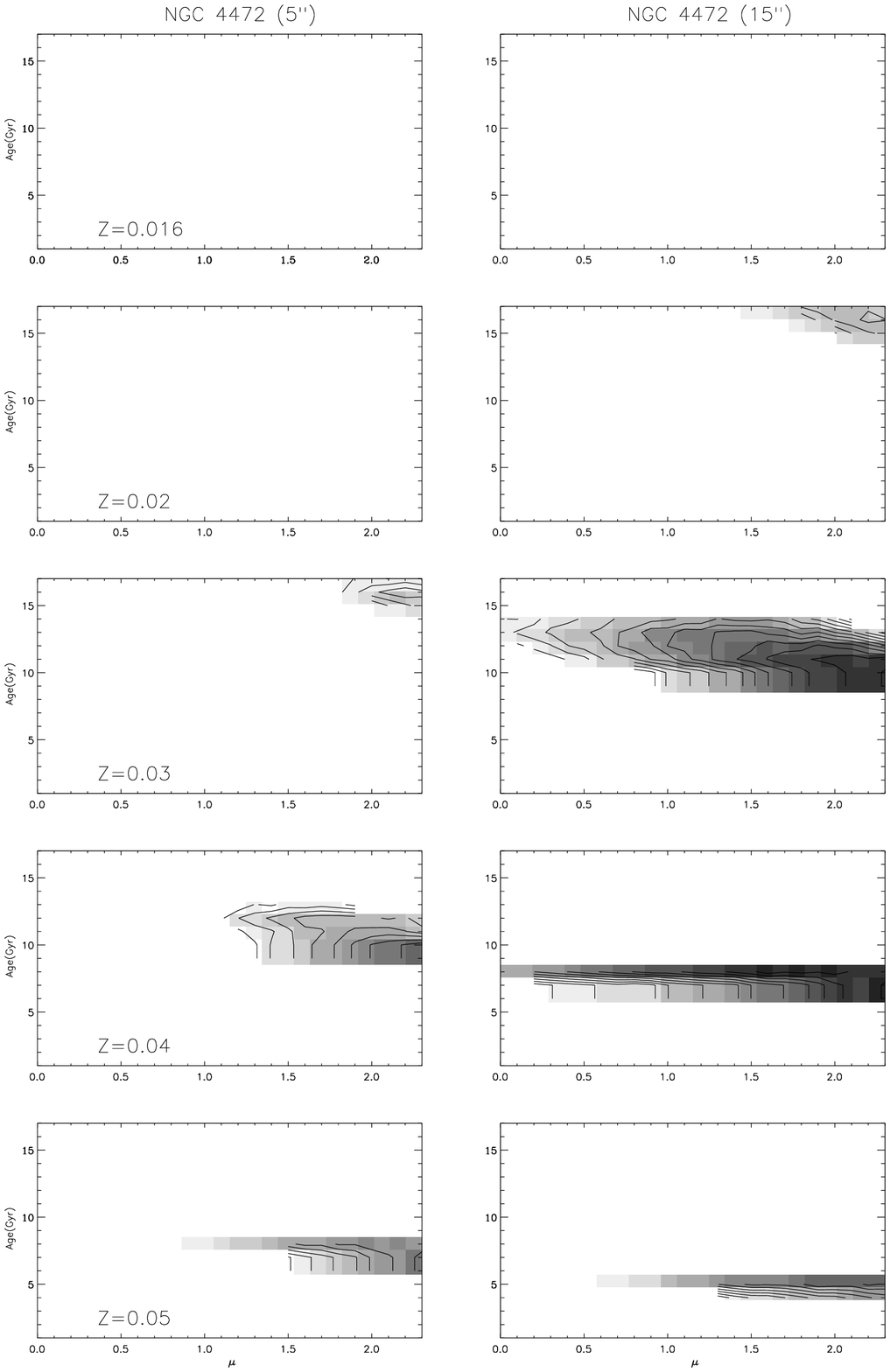}
\caption{The merit values obtained for NGC~4472 with our SSP models using a bimodal IMF.}
\end{figure}

\begin{figure}
\plotone{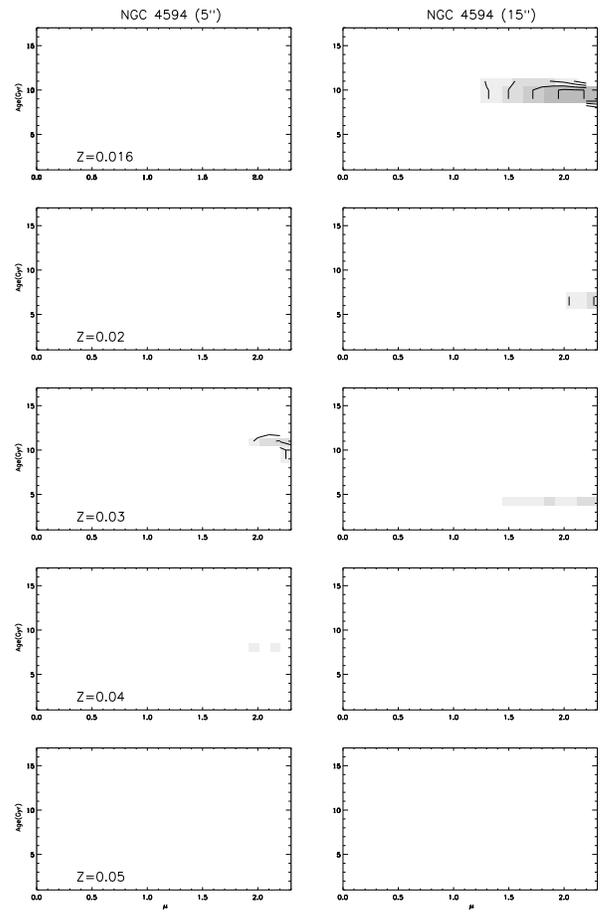}
\caption{The merit values obtained for NGC~4594 with our SSP models using a bimodal IMF.}
\end{figure}

\subsection{Fitting with the chemical evolutionary model}
Using V-K - Mg$_{2}$ and U-V - H$_{\beta}$ diagrams we found in Paper I that for a {\em closed-box}
approximation these metal rich galaxies cannot be fitted with a single 
IMF which is constant in time using the full chemical evolutionary model. 
The reason for this is the impossibility of producing a dominant old 
but metal-rich population in our observed galaxies. The same result was 
found in Casuso {\it et al.} (1996) on the basis of the Mg$_{2}$ index. To solve the problem, we proposed a
scenario invoking an IMF skewed towards high-mass stars during a short initial
period $t_{0}$ (smaller than 1~Gyr), followed by preferential low-mass star
formation during the remaining time. 
In that paper the favored level of metallicity reached after this short period
was around $\sim2~Z_{\odot}$ for fitting the inner regions of the three galaxies. We will use here only the
bimodal IMF, since we showed in Section~3.1.4 that this works slightly better than the unimodal IMF for a unique SSP model. This was also shown in Paper I for
the evolutionary model for a limited set of indices.  

\subsubsection{The best fits}
Once again we work with our Merit function to account for the whole set of
colors and line-strengths contained in the observed data. The parameter space
to be scanned is therefore bounded by the values proposed in Table~8 of Paper I
for the pairs ($\nu$,$\mu_{0}$) which drive the gas to higher than solar
metallicity at $t_{0}$. However since in the present paper we not only want to 
fit the three galaxies at 5\arcsec,  but also at 15\arcsec, it is also required to use the pairs of parameters which drive the metallicity to solar after the initial period, as in fact was shown
 in Section~3.1.6 using the SSP models. We scanned this initial period of time with values of 0.2, 0.5 and 1~Gyr. Finally we choose to scan the remaining IMF slope $\mu$
between 1.3 and 2.3, while the age of the galaxy is varied between 1 to 17~Gyr.

To illustrate our best fits, in Figs.~18 to 20, we have presented those merit figures
using the weights of Case A (see Section~3.1.6) which are lower than $2.5 \times
M_{max}$ (25.0) for $t_{0}$ equal to $0.2~Gyr$. For the sake of brevity we did
not plot here the cases for $t_{0}$ equal to $0.5$ and $1~Gyr$ because, even
if they show better fits than those in Figs.~14 to 17, they are in general slightly worse than
those obtained with $0.2~Gyr$. For this reason we concentrate on this latter 
value for the initial period. We point out that both the final metallicity as well as the average metallicity are now non free input parameters as in SSP 
models, because they depend on the selected $t_{0}$, $\nu$, $\mu_{0}$, $\mu$ and the age of the galaxy which determine the followed chemical evolution. We also point out that in these plots we selected the pairs ($\nu$,$\mu_{0}$) to drive the metallicity to $\sim2\times Z_{\odot}$ at $t_{0}$ except for the last two rows of merit diagrams of each figure in which the metallicity raises only to 
values around solar at $t_{0}$. However, different chemical evolutions could be achieved by varying $\mu$ and the age of the galaxy, 
driving the metallicity to values different from those obtained at $t_{0}$. For example, if we make the IMF slope low and the age high the metallicity continues raising (see for details Paper I).

A general trend which can be found looking at Figs.~18 to 20 is that acceptable fits for the three galaxies are in fact possible. Also one can notice that the grey levels are now darker than in Figs.~14 to 17, indicating
lower merits and hence better fits than the ones obtained with SSP models. 
We see that the prediction of high ages for this set of galaxies is now robust. 
In fact, ages above $\sim 10~Gyr$ are preferred to match the data. We also see
that to fit the three galaxies we require values of $\mu$ in the second phase 
always greater than $\sim 1.7$, indicating that after the first rapid formation
the formed stars were primarily of lower mass than in the solar neighborhood
($\mu=1.35$). Solutions involving $\mu<1.35$ for the second period are not
acceptable, because then we cannot stabilize the chemical evolution which continues raising the metallicity to higher values (even higher than Z=0.1).
All this shows that low-mass stars are very important
in this kind of systems as was pointed out by e.g. Faber \& French (1980). 

Another general conclusion is that we require the SFR coefficient ($\nu$) to be 
in the range $2.5\leq\nu\leq30\times10^{-4}Myr^{-1}$ (with low $t_{0}$) meaning
that our values are high compared to the solar neighborhood, estimated to be 
$\sim 1.92\times10^{-4}Myr^{-1}$ (Arimoto \& Yoshii 1986), but much lower than
the values proposed by these authors in their second paper (1987) to fit the
ellipticals ($\sim 96\times10^{-4}Myr^{-1}$). The reason for this difference is
that they stop the star formation after the occurrence of the galactic winds
i.e. after 1~Gyr of their model. In our study such high values do not emerge
because we do not stop the star formation (unless all the gas has been consumed,
precisely when using such high values of $\nu$ with normal or even 
higher IMF slope) and therefore unacceptable values for the metallicity for the remaining
gas are obtained. This happens especially when working at low IMF slopes. This
difference in input physics has important consequences to the fits obtained, 
since models which stop the
star formation after a short initial period of time only allow the formation of
stars with a mixture of metallicities but never with a mixture of ages. 
We also differ from Arimoto \& Yoshii (1987) in the prediction of the slope of
the IMF where they assume to be $\sim 1.0$ (slightly lower than Salpeter). To 
compare our results with theirs we have to look at the results obtained with
the unimodal IMF. The prediction obtained by ourselves in Paper I (as can be 
deduced from Fig.~20 of that paper) is around $\sim1.7$, therefore higher than 
Salpeter and than their estimate. We attribute this difference to the fact that 
they need this combination of lower than Salpeter IMF and very high star 
formation rates to yield the required high metallicities in their period previous 
to the occurrence of the galactic wind (1~Gyr), while we obtain this in a 
shorter time by means of lower IMF slopes than they require. Bressan et al. 
(1994) also follow the galactic-wind scheme, without changing the Salpeter IMF,
but requiring a shorter period of time before the appearence of the galactic 
winds. These short periods of initial different behaviour are in better 
agreement with our best fits.

Now turning our attention to the differences within galaxies and between them, we see that NGC~3379 and the bulge of Sombrero galaxy appear to lie in the same regions of ($\mu$,age) parameter space for a given ($\nu$,$\mu_{0}$). We also see that, for the same set of parameters $\nu$ and $\mu_{0}$, NGC~4472 at 5\arcsec~  gives better fits for ages which are greater by $\sim 3~Gyr$ than
 the ones obtained for the other two galaxies at the same distances from their
 centers. However, by looking at the results obtained with the SSP models and the metallicities obtained in the best fits 
(given in Table~7) we are biased to think that after the
initial period, $t_{0}$, NGC~3379 and NGC~4594 reached lower metallicities than NGC~4472, as also was shown in Paper I. This also holds when varying the value of the IMF slope for the remaining time.
The full chemo-evolutionary models also produce the same conclusion as the SSP models when fitting the observed gradients, i.e. the metallicities of the outer regions are lower for NGC~3379 and NGC~4594 but not noticeably lower for NGC~4472, probably because the latter is too large and therefore the radial separation of the observed points does not represent an important fraction of 
its effective radius. In fact in Figs.~18 and 20 we see that the outer regions of NGC~3379 and NGC~4594 are only fitted in an acceptable way for the last two rows of merit figures, where the metallicity does not raise above solar at $t_{0}$. For more details about the final and average metallicities see Table~8, where we summarize the most representative fits. We also point out that in this table comparing the B-V and V-K colors of NGC~3379 and Sombrero galaxy 
we are inclined to think that
the latter could be affected by a small amount of dust, since its colors are 
too red while the lines seem to be well fitted. We think it is likely that a reddening correction of E(B-V)$\sim$0.05 should be applied. However, the improvement in our fit will not be very large, given the size of the error-bars.

Finally, in Figs.~21 and 22 we show the chemical and fractional gas mass evolution as 
well as the distribution of the {\em live} stars as a funcion of metallicity, 
age and temperature for the two selected positions of NGC~3379. We see that at 5\arcsec~  the stars mainly 
have metallicities larger than solar  (weighting in the V band) but with approximately $20\%$ of the 
light coming from stars with metallicity of solar or lower. 
However at 15\arcsec~  we see that most of the stars (around $90\%$) have solar metallicity. From the diagrams giving the predicted distributions of
stars as functions of time one can also notice that the bulk of 
the stars were formed at an early-stage of the evolution of the galaxy, 
at ages lower than some $\sim 1.5~Gyr$. In general, for the three galaxies 
we obtain that the contribution to the light in the U band is around  
$\sim30\%$ for giants and $\sim70\%$ for dwarfs, in the V band it is 
$\sim50\%$ for giants and $\sim50\%$ for dwarfs, while in the K band the 
proportions are $\sim75\%$ for giants and $\sim25\%$ for dwarfs respectively.

\begin{figure}
\plotone{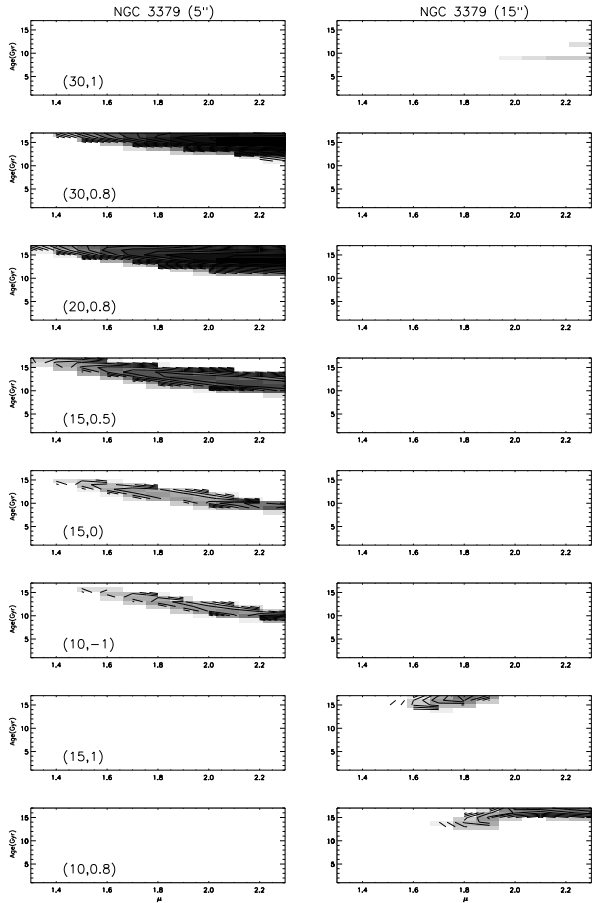}
\caption{The merit function values in ($\mu$,Age) space obtained with our full chemo-evolutionary population synthesis model using a bimodal IMF. The initial period in which the IMF was skewed towards massive stars was 0.2~Gyr. 
Models for various parameters ($\nu$,$\mu_{0}$) (taken according to Table~8
of Paper I) are selected. 
The pairs ($\nu$,$\mu_{0}$) of the first 6 rows of merit diagrams drive the metallicity to $\sim 2\times Z_{\odot}$ at $t_{0}$, while the pairs corresponding to the last two rows drive the metallicity to solar at $t_{0}$.
The contours and the grey-scale have been defined using the same criterion as in Figs.~14 to 17. Notice that the fits are better since the 
greyscales are now darker and occupy wider regions of the parameter space.}
\end{figure}

\begin{figure}
\plotone{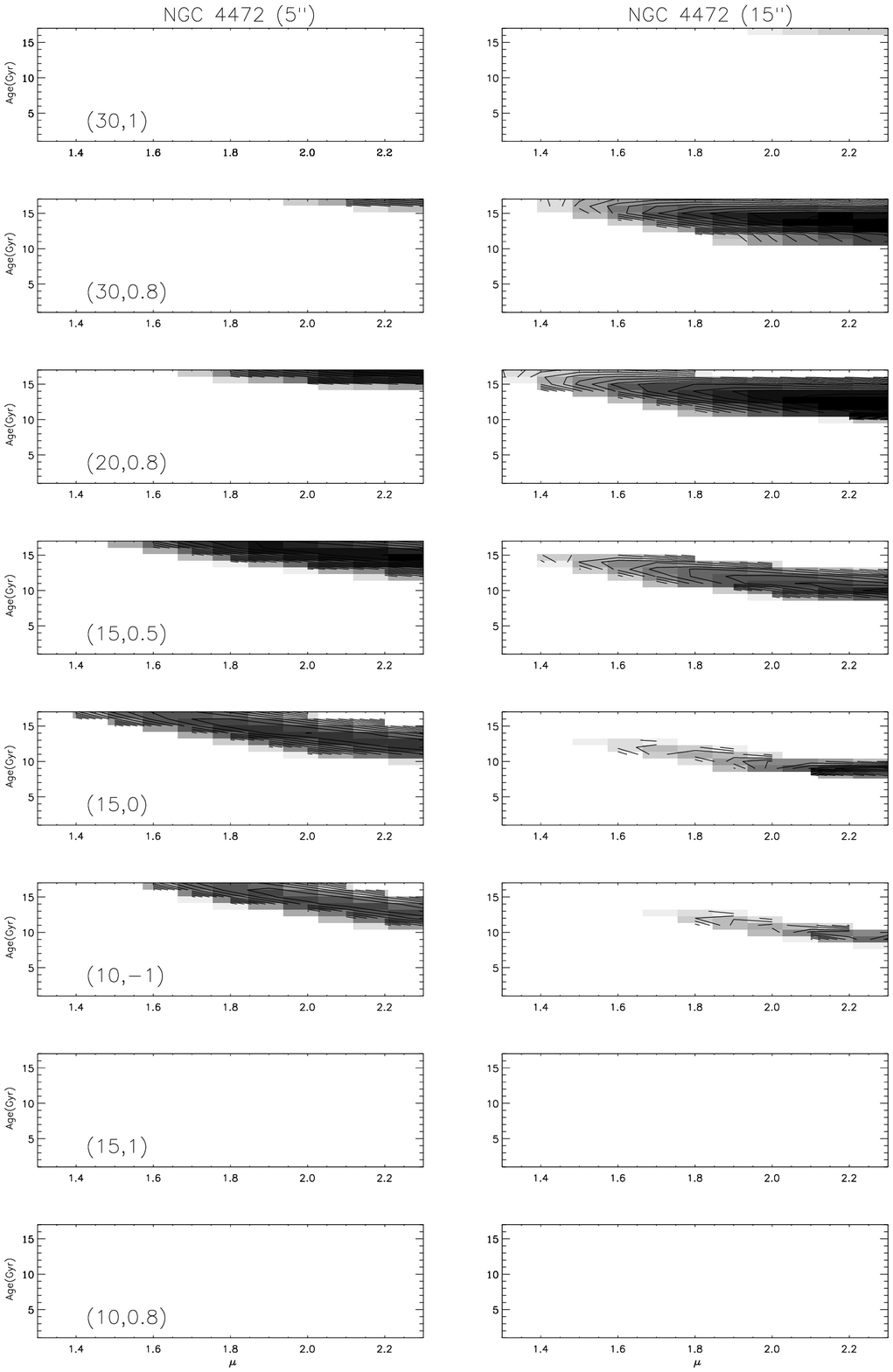}
\caption{The merit function values for NGC~4472 obtained with our full chemo-evolutionary population synthesis model using a bimodal IMF.}
\end{figure}

\begin{figure}
\plotone{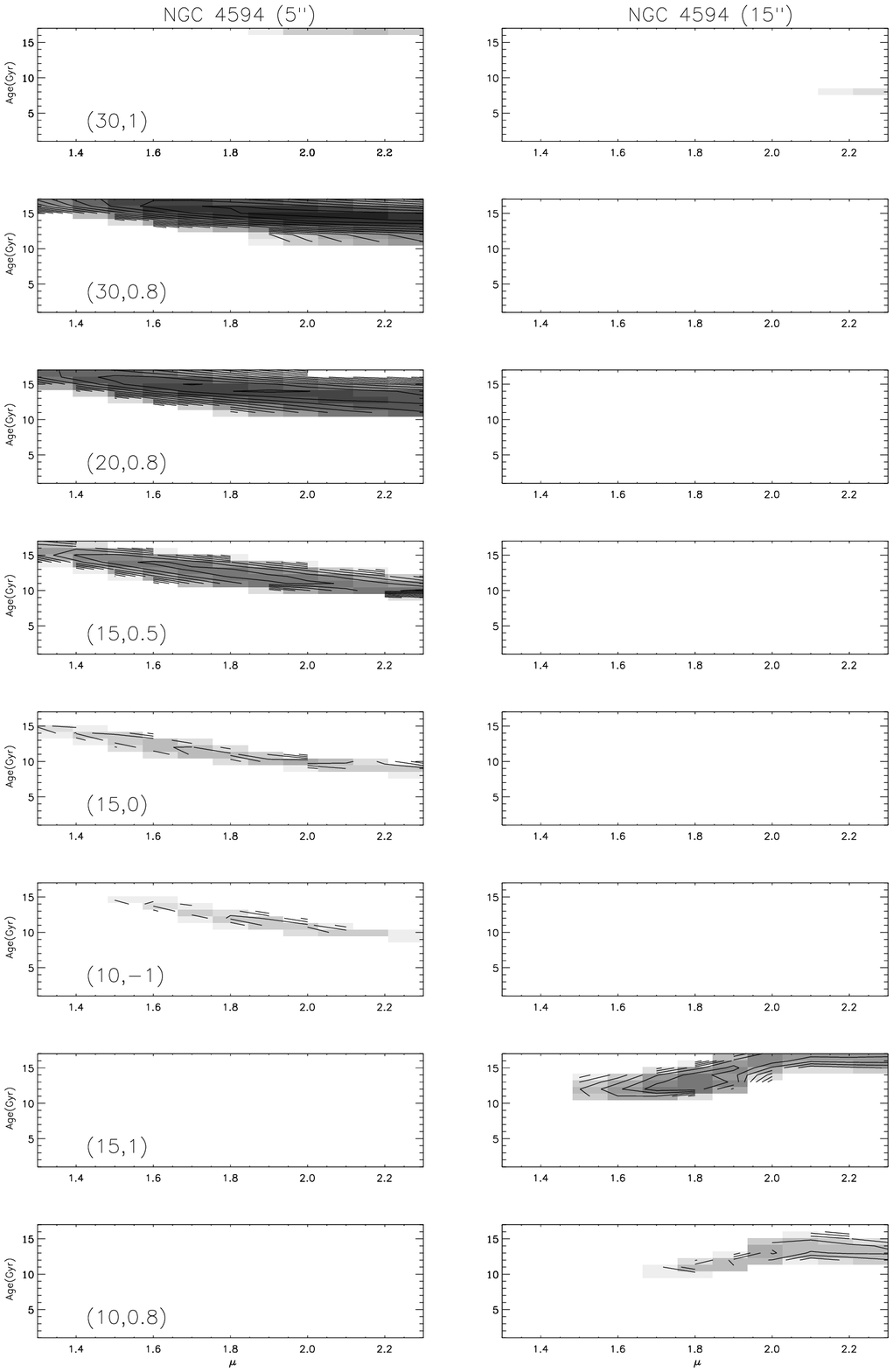}
\caption{The merit function values for NGC~4594 obtained with our full chemo-evolutionary population synthesis model using a bimodal IMF.}
\end{figure}

\begin{figure}
\plotone{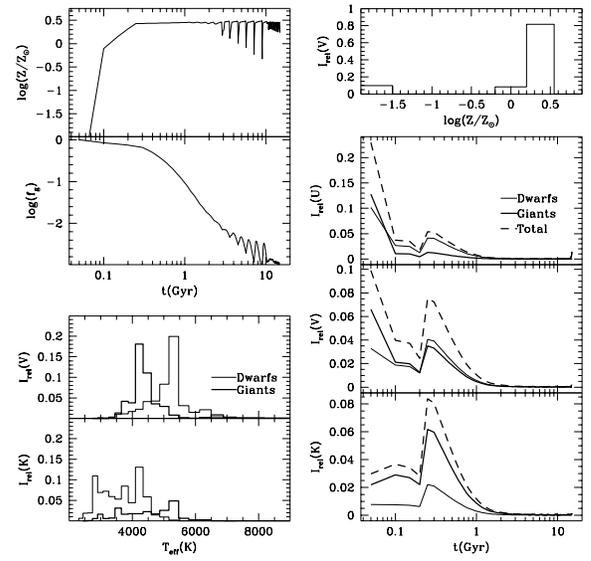}
\caption{The chemical and fractional gas mass evolution and the predicted 
distributions of the live stars as a function of the metallicity, time and
temperature for different broad band filters in a representative solution for
NGC~3379 at 5\arcsec. This fit was obtained for $t_{0}=0.2~Gyr$, $\mu_{0}=0.8$, $\nu=30$,
$\mu=2.3$ and assuming an age of 15~Gyr. The quantity $I_{rel}$ represents the
fraction of the total luminosity.}
\end{figure}

\begin{figure}
\plotone{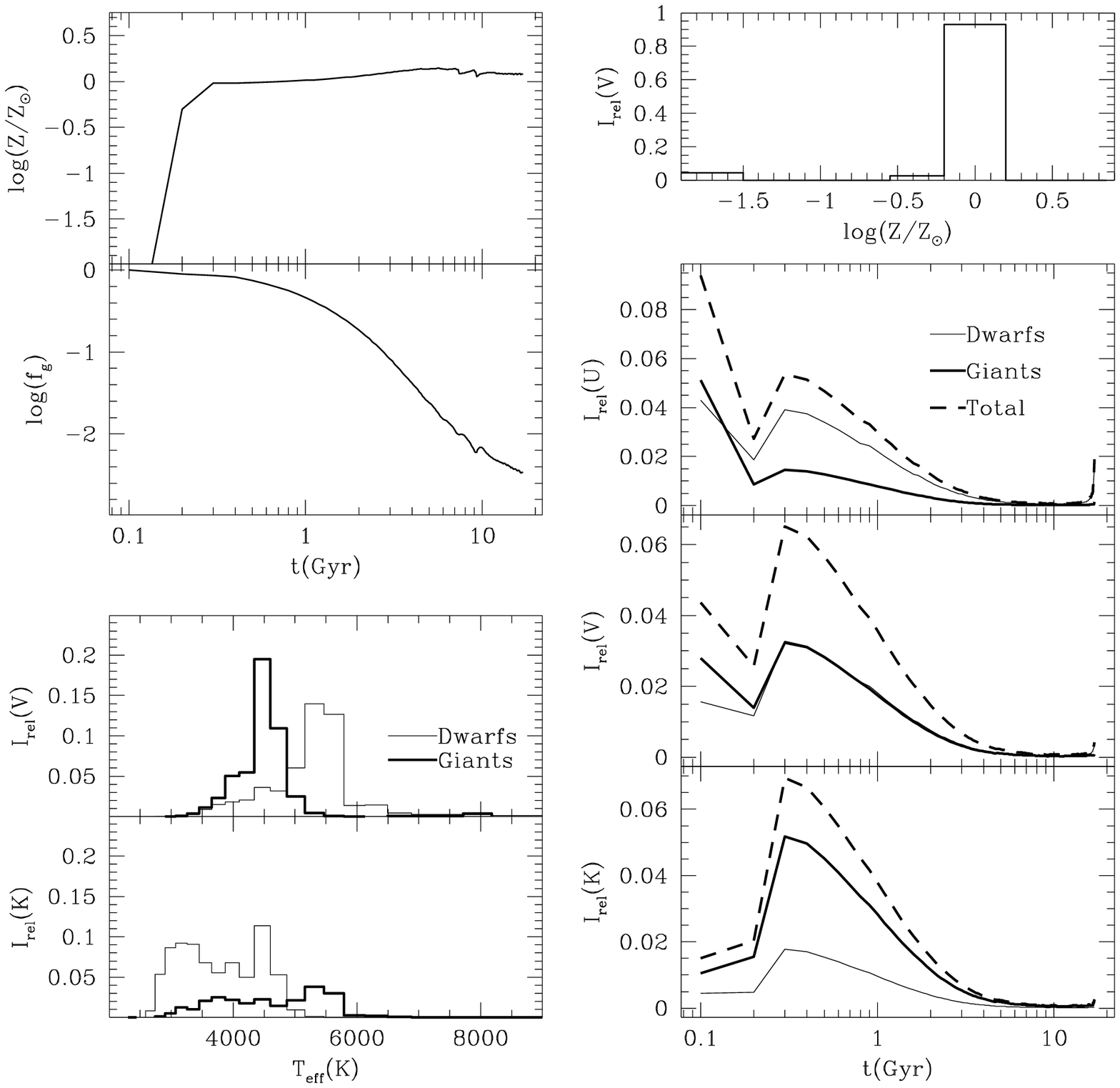}
\caption{The same plots as in Fig.~21 for a representative solution for NGC~3379 at 15\arcsec. This fit was obtained for $t_{0}=0.2~Gyr$, $\mu_{0}=0.8$, $\nu=10$, $\mu=2.3$ and an age of 17~Gyr. }
\end{figure}

\subsection{SSP models versus chemical evolutionary models}
We find that using the evolutionary model most of the stars were formed in the 
early stages ($< 1.5~Gyr$) of the galactic evolution, as seen in Figs.~21 and 22. 
Therefore, it is not surprising that qualitatively the fits from the evolutionary model are not very different from those of the single-age, single-metallicity models. Somewhat better fits are obtained however, indicating that these galaxies probably have stars with a mixture of metallicities, as has been found also in our Galactic bulge (Rich 1988). 

It is also easy to understand that the evolutionary model gives somewhat larger
ages, since the stars of lower-metallicities have to be compensated by giving
the whole population a greater age.
 
\section{Conclusions}
We have obtained high quality observations of almost the whole set of
line-indices of the extended Lick-system for three representative 
early-type galaxies, and have applied a new spectrophotometric 
chemo-evolutionary population synthesis model developed by ourselves
 in a previous paper (Paper I).  

We can make models which give good fits in all the colors and many 
of the most important line-indices. These fits however cannot 
synthesize quantitatively a number of lines primarily from Fe and 
Ca. We find that 6 independent Fe lines are too weak compared 
to lines of all other elements indicating that the 
iron-abundance is anomalous and deficient in the 
radial range of the galaxies that we studied.
This implies that the global metallicity inferred 
must depend on whether we use magnesium or iron lines as the prime 
indicators. Invoking $\alpha$-enhancement one can obtain better 
fits for the iron lines, but other features such as NaD then become 
worse if we follow the abundance ratios given in Weiss {\it et al} 
(1995). Finally we find that the Ca4227 is much fainter than 
predicted by the models.

In general we find that the three galaxies require metallicities
higher than solar for the inner regions while the ages are older 
than 10~Gyr, and the observed radial gradients are due to 
metallicity decreasing outward. We also find that NGC~4472 is more 
metal-rich, than the other two galaxies.

To fit this set of galaxies with the full chemical evolutionary 
population 
synthesis model we used the variable IMF scenario (defined in 
Paper I) which invokes an IMF skewed towards high-mass stars in 
the begining, during a short period of time (smaller than 1~Gyr.), 
and towards low-mass stars later for the remaining time. The best 
fits indicate that dwarfs contribute $\sim 70\%$ to the U band, 
$\sim 50\%$ in V band and $\sim 25\%$ in K band.

We find that slightly better fits are being obtained with the chemical evolutionary model than with the single-age, single-metallicity model, justifying the extra complications. However, since the predicted spread in metallicities is not very large, and since the bulk of the stars were formed at the very early stages of the 
galactic evolution (at age lower than $\sim1.5~Gyr$) we conclude 
that the single-age single-metallicity stellar population models 
offer reasonable first order fits to this kind of stellar systems, 
especially if one wishes to avoid comprobational complexity.

This study shows that it would be useful to extend the present 
analysis to include other features at shorter wavelengths in the 
UV region such as the indices of Rose (1994), and to the near-IR 
with indices such as Na$_{I}$ at $8190~{\rm \AA}$, the Ca$_{II}$
triplet and the CO or H$_{2}$O features. To understand the stellar
populations of the early-type galaxies and to e.g. disentangle age 
and metallicity (Jones \& Worthey 1995), Bressan {\it et al.} 1995) 
it will be important to introduce as many constraints as possible, 
by observing the galaxies in many calibrated absorption lines.

The observational data presented in this paper are available on 
the AAS CD-ROM, and from the WWW-homepages of the authors. 

\acknowledgments
We are indebted to the referee, J. Gonzalez, who have done a very thorough
work on our first version as a result of which the present paper have
been greatly improved.
We are grateful to J. Gorgas for providing us with a set of observational
results before publication. We also thanks A. Weiss for his comments. 
This work was partially supported by grants
PB91-0510 and PB94-1107 of the Spanish DGICYT.

\onecolumn
\begin{figure*}
\plotone{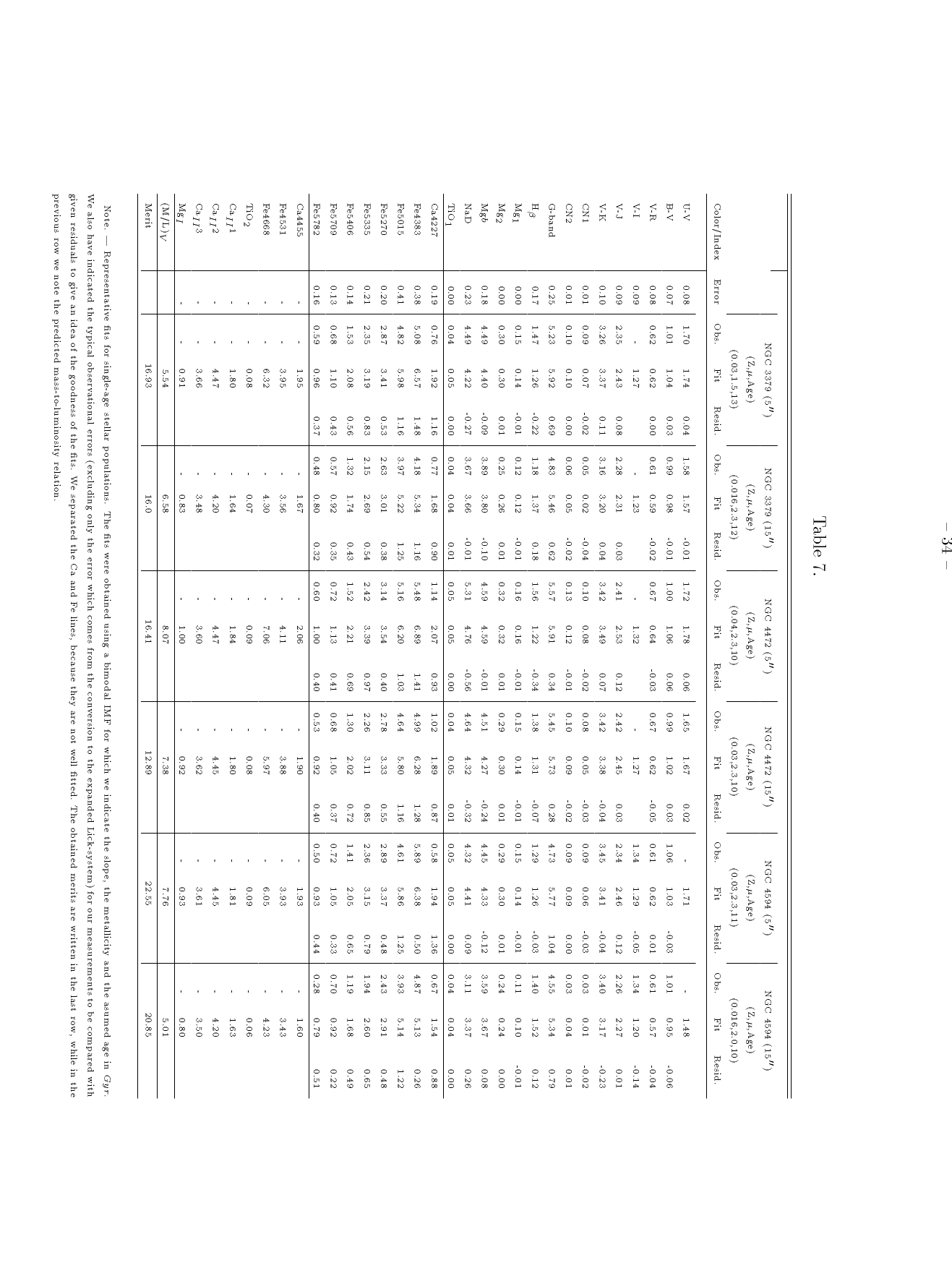}
\end{figure*}
\begin{figure*}
\plotone{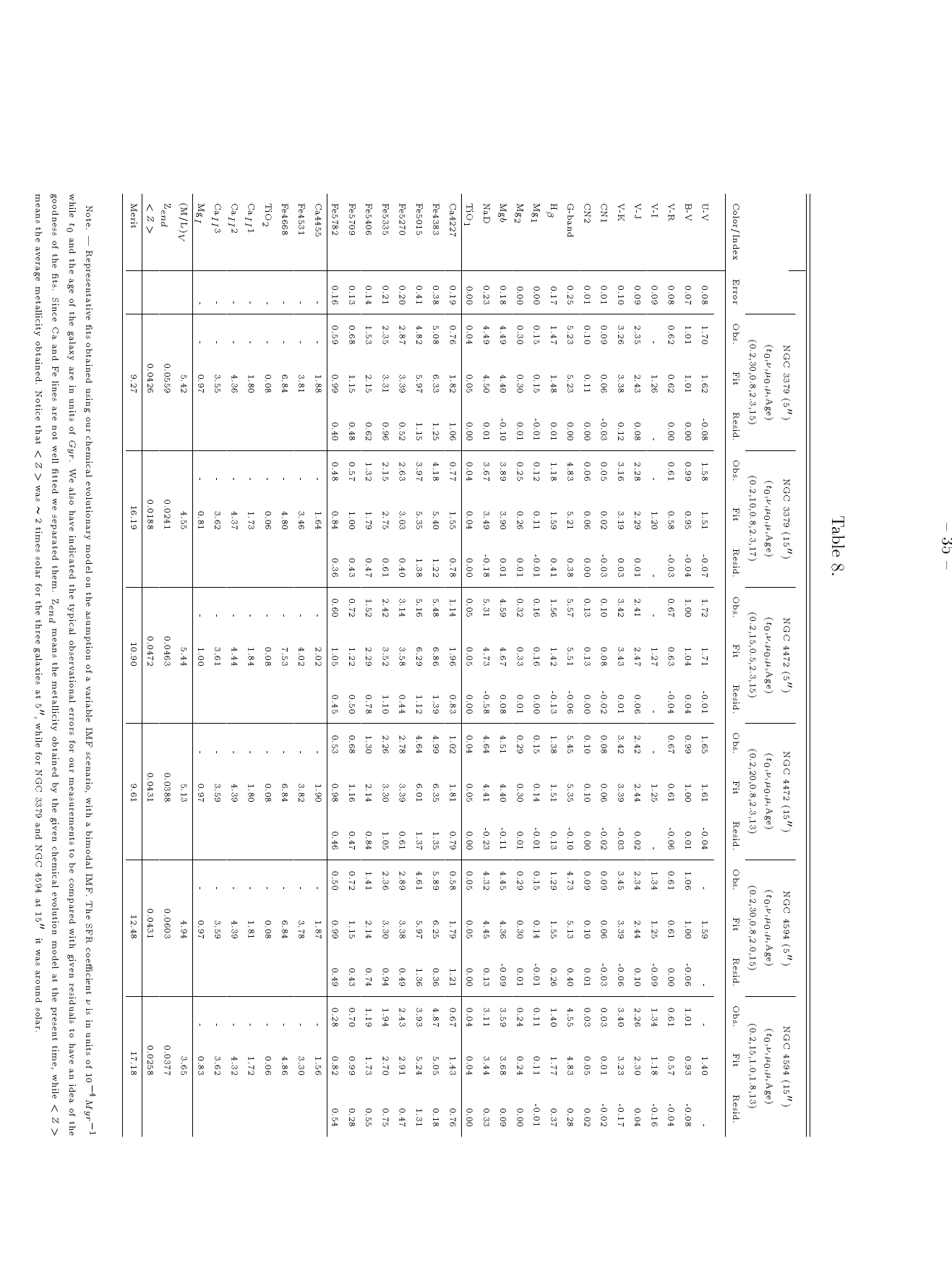}
\end{figure*}

\end{document}